\documentclass{aa}  

\usepackage{natbib}
\bibpunct{(}{)}{;}{a}{}{,} % to follow the A&A style
\usepackage{graphicx}   % Including figure files
\usepackage{txfonts}

\usepackage{amssymb}    % Extra maths symbols
\usepackage{import}
\usepackage{subfigure}
\usepackage{multirow}
\usepackage{multicol}
\usepackage{array,tabularx}
\usepackage{graphicx}
\usepackage{xcolor}
\usepackage{newtxtext,newtxmath}
\usepackage{ulem}
\usepackage{float}

\newcommand{\bvec}[1]{\mathbf{#1}}
\newcommand{\sa}[1]{{\color{black}#1}}

\begin{document}

   \title{A numerical study on the role of instabilities on multi-wavelength emission signatures of blazar jets}

   %\subtitle{Simulating the impact of Instabilities on variable Blazar jets}

   \author{Sriyasriti Acharya\inst{1}\fnmsep\thanks{E-mail: sriya.acharya@gmail.com}
          \and
          Bhargav Vaidya\inst{1}  \and Indu Kalpa Dihingia\inst{1,2,3} \and Sushmita Agarwal\inst{1} \and Amit Shukla\inst{1} 
          }

   \institute{Department of Astronomy, Astrophysics and Space Engineering, Indian Institute of Technology Indore, Khandwa Road, Simrol, Indore 453552, India \and Department of Physics, Indian Institute of Science, Bengaluru, Karnataka 560012, India \and Tsung-Dao Lee Institute, Shanghai Jiao-Tong University, 520 Shengrong Road, Shanghai, 201210 People's Republic of China}

   \date{Received XXX; accepted XXX}

  \abstract
  {Blazars, \sa{a class of active galaxies whose jets are} relativistic and collimated flows of plasma directed along the line of sight and are prone to a slew of magneto-
hydrodynamic (MHD) instabilities. These jets show characteristic multi-wavelength and multi-timescale variability.}
   {We aim to study the interplay of radiation and particle acceleration processes in regulating the multi-band emission and variability
signatures from blazars. In particular, the goal is to decipher the impact of shocks arising due to MHD instabilities in driving the long
term variable emission signatures from blazars.}
   {In this regard, we have performed RMHD simulations of a representative section of blazar jet. The jet is evolved using a hybrid Eulerian-Lagrangian framework to account for radiative losses due to synchrotron process and particle acceleration due to shocks. Additionally, we have incorporated and validated radiative losses due to the external Compton (EC) process that are relevant for blazars. We have further compared the effects of different radiation mechanisms through numerical simulation of 2D slab jet as a validation test. Finally, we have carried out a parametric study to quantify the effect of magnetic fields and external radiation field characteristics by performing 3D simulations of a plasma column. The synthetic light curves and spectral energy distribution (SEDs) are analysed to qualitatively understand the impact of instability driven shocks.}
   {\sa{We observe that shocks produced with the evolution of instabilities give rise to flaring signatures in the high energy band. The impact of such shocks is also evident from the instantaneous flattening of the synchrotron component of the SEDs.} At later stages, we observe the transition in X-ray emission from the synchrotron process to that dominated by EC. The inclusion of the EC process also gives rise to $\gamma$-ray emission and shows signatures of mild Compton dominance as typically seen in Low Synchrotron Peaked blazars.}
   {}

  \titlerunning{Simulating the impact of instabilities on variable blazar jets}
  \authorrunning{S. Acharya et al.}

   \maketitle
%
%-------------------------------------------------------------------

\section{Introduction}

Blazars belong to the radio-loud subclass of Active Galactic Nuclei (AGN) \citep{Blandford_1978_Conf_proc, Blandford_1979, urry_1995, Blandford_review_2019, Hardcastle_2020}, with the jet directed along the line of sight of the observer \citep{Dondi_1995}. As a result, the jet radiation is enormously intensified due to the relativistic boosting effect, \sa{appearing as} one of the most \sa{dominant sources for extra-galactic $\gamma$-ray sky \citep{Paliya_2019, Hovatta_2019_review, Bhatta_2022_review}.} The typical two hump structure in its spectral energy distribution (SED) is mainly dominated by two broad non-thermal radiation components \citep{Urry_1999}. The low energy hump is attributed to the synchrotron emission from the relativistic electrons that typically extend from radio \sa{to UV or X-ray \citep{Bottcher_2007_review, Bottcher_2010, Meyer_2012}}. Often, the presence of different particle energization processes may be accountable for the observed broadening of the low energy hump beyond the X-ray band \citep{Kirk_1998}. The source of the high energy hump is believed to be due to the inverse Compton scattering of low energy photons and extends all the way to the high-energy end of $\gamma$-rays \sa{\citep{Liodakis_2018}}. Furthermore, the production of neutrino emission \citep{IceCube_2013} from the blazars suggests pion decay, in addition to proton synchrotron radiation, as the processes responsible for the high energy emission \citep{Mannheim_1993, Mucke_2001_proton_sync, Petropoulou_2015_hadronic}.

Based on the optical spectra, the unification scheme of radio-loud AGN classified blazars into two broad categories: Flat Spectrum Radio Quasars (FSRQs) and BL Lac objects \sa{\citep{Stickel_1991, Stocke_1991, urry_1995}}.  
The Compton dominance is larger in FSRQs due to the presence of external photon sources such as accretion disk, broad and narrow line regions (BLR $\&$ NLR respectively), torus, etc. In these sources, external Compton (EC) \citep{Begelman_1987, Dermer_1992, Dermer_1993_EC, 1994_Sikora_EC, Kataoka_1999, Madejski_1999, Bazejowski_2000, Ghisellini_2009} is the dominant mechanism responsible for the second hump. In certain cases, isotropically present CMB photons are also scattered by the relativistic electrons, giving rise to the IC-CMB process \citep{Bottcher_2008_IC-CMB, Meyer_2015, Zacharias_2016}.
However, the BL Lacs are low power sources and exhibit smaller Compton dominance, suggesting an insignificant contribution from EC. In these sources, synchrotron self-Compton (SSC) \citep{Marscher_1985_SSC, Bloom_1996} plays a vital role as a possible mechanism for the second hump. The Compton dominance also decreases as the peak frequency of the lower energy hump shifts towards a higher range \citep{Finke_2013, Prandini_2022}. Such features are explained based on the strength of radiative cooling suffered by the emitting electrons for different sources \citep{Ghisellini_1998}. In certain blazars, the SED shows unusual characteristics. For example, the multi-wavelength observations of AO 0235+164 show a triple hump structure during its flaring state \citep{Ackermann_2012}. Furthermore, flattening of synchrotron spectra at high energy range is also observed that results in change in the slope of X-ray component in the valley of the SED \citep{Bottcher_2003_BLLacertae, Sahakyan_2022}. However, the origin of such changes in SED from its classical behaviour is still unresolved in spite of several propositions that have been put together to explain such phenomena.

Blazar emission also exhibits multi-timescale variability, with the occasional appearance of quasi-periodicity. Several scenarios have been addressed before as a source of a possible explanation of such observed variability timescales. For example, propagation of blobs of plasma through helical magnetic fields \citep{Marscher_2008}, presence of shocks in the jets \citep{Valtaoja_1992A, Valtaoja_1992B, Turler_2000}, \citep{Larionov_2013}, geometrical effects related to the viewing angle of the observer with respect to the emission zone \citep{Villata_1998, Raiteri_2017}, magnetic reconnection \citep{Ghisellini_2008_FastTeV, Giannios_2009, Giannios_2010, Narayan_2012_JetinJet, Giannios_2013, Shukla_2020}, presence of binary black hole systems \citep{Begelman_1980}, \citep{Sillanpaa_1988, Gupta_2019}, etc.
Recent studies in jets have shown that the optical polarisation signatures can be highly variable and correlated with the high activity state \citep{Abdo_2010_polarization, Kiehlmann_2016}.

Jet instabilities play a major role in governing the observed signatures of blazar emission. For example, the fluctuations observed in the polarisation angle of blazar jets \citep{Zhang_2017} and the quasi-periodic nature of the blazar emission may have a kink origin \citep{Dong_2020}. Additionally, the long term variability could also be explained through a kink driven helical jet model \citep{Acharya_2021}.
Studying the temporal variation of flux provides a broad picture of the source, whereas modelling multi-wavelength spectra is required to understand the substantial and extreme physical conditions within the emission region. In addition, it is also crucial to understand different particle energization mechanisms (such as shock acceleration and magnetic reconnection) and their impact on the light curve and broad-band spectra. In the case of highly magnetised environments, it was demonstrated that the kink instability generated current sheets might act as possible particle acceleration sites \citep{Bodo_2021, Kadowaki_2021}. Some studies also proposed that the blazar flares may be powered by internal shock model \citep{Marscher_1985_SSC, Mimica_2004, Bottcher_Dermer_2010, Moderski_2003, Joshi_2011}. \cite{Hovatta_2008} investigated the long-term radio variability of a sample of AGN flares, and their findings appear to be consistent with the shock-in-jet theory.
\cite{Fichet_2022} have also showed that a strong interaction between the standing shock and moving shock may lead to the generation of flares in the light curve.
The above-mentioned studies include several simplifications such as steady state single zone emission modelling and incorporating simplified or limited high energy emission mechanisms. \cite{Moderski_2003} presented a code that simulates light-curves and spectra of blazars during flares by considering a single blob as the emitting region. There have also been approaches to study blazar jet emission by performing 3D numerical simulations with fixed spectra (not accounting for all particle acceleration processes) \citep{Dong_2020, Acharya_2021, Kadowaki_2021, Fichet_2022}.
To improve upon these simplifications, we extend our previous work of \cite{Acharya_2021} by incorporating consistent emission mechanisms with the hybrid framework of the \texttt{PLUTO} code \citep{Vaidya_2018, Mukherjee_2021}.

In particular, we have included a time-dependent multi-zone emission model with suitable radiative and particle acceleration mechanisms (particularly due to shocks).
Several numerical studies have combined the macroscopic fluid flow with the microscopic Fokker Planck solver to study shock acceleration. For example,  \citep{Achterberg_1992, Marcowith_1999, Wolff_2015} have adopted a stochastic differential solver to connect these scales. Alternatively, hybrid framework where evolution of a non-thermal electron is followed with the dynamics of the fluid is adopted in several works to study effect of particle acceleration \citep{Micono_1999, Tregillis_2001, Mimica_2012, Vaidya_2018, Fromm_2019, Winner_2019, Huber_2021}.

Our principal aim is to investigate the impact of shock acceleration on the broad-band emission characteristics of blazar jets in the presence of various magneto-hydrodynamic (MHD) instabilities.
The contents of this paper are divided into two parts. Initially, we focus on the numerical implementation of inverse Compton scattering of external photon fields and understand the contribution of different parameters on the particle spectra and the emissivity. Subsequently, we adopt this implementation to study emission signatures associated with 2D relativistic slab-jet toy model and 3D cylindrical plasma column with different magnetisation values. 

The paper is structured as follows:
In section \ref{sec:IC_method}, we describe the numerical implementation of the external Compton process for a mono-directional source of seed photons. This section includes the calculation of energy loss rate and external Compton scattering emissivity. Section \ref{sec:slabjet} is devoted to the results obtained from a 2D relativistic slab jet simulation along with the numerical setup and emission modelling approach. In section \ref{sec:plasma_column}, we explain the multi-wavelength nature of a relativistic plasma column that has undergone kink instability and show a comparative analysis among different parameters responsible for the highly energetic external Compton process. Finally, we discuss all the results and summarise our current findings in section \ref{sec:discussion} and \ref{sec:summary} respectively.

\section{Implementation of external Compton mechanism}\label{sec:IC_method}

We have adopted the \texttt{PLUTO} code \citep{Mignone_2007_PLUTO} for all studies in the current work. We have specifically used the hybrid framework in the \texttt{PLUTO} code \citep{Vaidya_2018, Mukherjee_2021}.
This framework is developed to model the non-thermal spectral signatures of the macro-particles in the relativistic MHD flow.
The macro-particles are assumed to be an ensemble of leptons with a finite energy distribution. Based on the fluid conditions, the spectral distribution of each particle is updated with time. This enables us to account for several physical processes and emission mechanisms, such as synchrotron and inverse Compton (IC) scattering. The current framework already has a numerical scheme to account for energy losses due to synchrotron and IC-CMB (IC scattering of CMB photons) processes. In this work, we have additionally incorporated the IC scattering process, where the origin of the target photon field considered for the IC scattering is external to the jet, such as the BLR region, accretion disc, etc. Such an emission mechanism is typically known as the external Compton (EC) process. In particular, we have incorporated the EC process considering a mono-directional photon field and all the formalisms provided in this regard are taken from \cite{Khangulyan_2014}.

\subsection{Calculation of loss rate}

A semi-analytical approach is considered by \cite{Vaidya_2018} for the evolution of the spectral distribution of the macro-particles using a Lagrangian scheme. For this purpose, they solve a characteristic equation to calculate the energy loss rate due to the above-mentioned physical processes and emission mechanisms. We extend the energy loss rate equation with an additional term that accounts for the loss due to the EC process,

\begin{equation}
    \frac{dE}{d\tau_{\rm p}} = -c_{1}(\tau_{\rm p})E - c_{2}(\tau_{\rm p}) E^{2} - c_{3} E f(E) \equiv \dot E, \label{eq:characteristic}
\end{equation}
where the first term in Equation \ref{eq:characteristic} represents the loss due to adiabatic expansion, the second term represents the loss due to both synchrotron and IC-CMB (under the Thomson limit), and the third term corresponds to the loss due to the EC process. $\tau_{\rm p}$ is the proper time measured in the jet co-moving frame. The constants associated with all these loss terms are given below,

\begin{align}\label{eq:loss_constants}
    \begin{split}
        c_{1} & = \frac{\nabla_{\rm \mu} \, u^{\rm \mu}}{3}  \\
        c_{2} & = \frac{4 \sigma_{\rm T} c \beta^{2}}{3 m_{e}^{2} c^{4}}[U_{\rm B} + U_{\rm rad}(E_{\rm ph})] \\
        c_{3} & = \frac{2r_{0}^{2} k_{\rm B}^3 T_{\rm cmv}^3 \kappa_{\rm cmv}}{\pi c^2 \hbar^3}
    \end{split}
\end{align}
In the above equations, c$_{1}$ can be calculated from the mass conservation equation as $\frac{\nabla_{\rm \mu}\, u^{\rm \mu}}{3}$ = $\frac{1}{3\rho} \frac{d\rho}{d\tau_{\rm p}}$ %= $\frac{2}{3} \frac{\rho - \rho_{\rm old}}{(\rho_{\rm old} + \rho)\Delta \tau_{\rm p}}$ 
with u$^{\rm \mu}$ being the bulk four-velocity and $\rho$ is the density of the fluid. 
Here, $\beta$ and m$_{\rm e}$ are the velocity (in the units of speed of light, $c$) and mass of the electrons, respectively, $\sigma_{\rm T}$ is the Thomson cross-section. The quantities U$_{\rm B}$ and U$_{\rm rad}$ are magnetic and the radiation ﬁeld energy densities, respectively, and E$_{\rm ph}$ is the energy of the incident CMB photon, r$_{0}$ is the electron classical radius. $\kappa_{\rm cmv}$ and T$_{\rm cmv}$ are the dilution factor and temperature of the photon field in the comoving frame of the jet. Boltzmann constant and Planck's constant are denoted in usual manner as $k_{\rm B}$ and $\hbar$ = h/2$\pi$ respectively. It is also important to take notice of the fact that the temperature and the dilution factor should be corrected for the relativistic bulk motion of the emitting region in the following way \citep{Khangulyan_2014}:
\begin{equation}
    T_{\rm cmv} = \frac{T}{D_{\rm PH}}, \quad \kappa_{\rm cmv} = D_{\rm PH}^2 \kappa,
\end{equation}
where T and $\kappa$ represent the photon temperature and dilution factor as measured in the lab frame. These quantities are set as input parameters in our model. $D_{\rm PH}$ = [$\Gamma$(1 - $\beta\cos\chi$)]$^{-1}$ is the Doppler factor between the source of seed photons and the EC emitting (jet) region, where $\chi$ is the angle made by the velocity vector of the electron with the photon's direction. For the present work, we have considered mono-directional photon field such that it lies beneath the EC emitting (jet) region and along the direction of jet motion (such that $\chi \sim 0$).
\sa{Inverse Compton scattering is therefore implemented by assuming a tail-on collision where the electron's velocity vector makes a zero degree angle with respect to the direction of the photon.}. 

\begin{figure}
    \centering
    \includegraphics[scale=0.44]{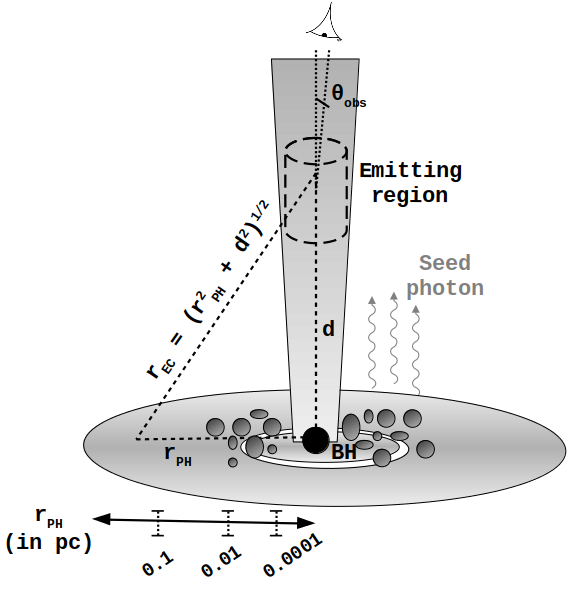}
    \caption{Cartoon representation of emitting region and external photon fields \sa{with seed photon's direction}. The size and distance of different components shown are not to scale.}
    \label{fig:EC_PH_cartoon}
\end{figure}

The dilution factor for a mono-directional photon field can be approximately represented as the following:
\begin{equation}
\kappa = \frac{\Delta \Omega}{4\pi},
\end{equation}
where $\Delta \Omega$ is the solid angle of the target photon field as observed from the EC emitting region. In our work, $\Delta \Omega$ $\ll$ 1 since the emitting region is considered tens of parsec away from the external photon field and, therefore, can be further expressed as below:
\begin{equation}\label{eq:kappa}
\kappa = \left(\frac{r_{\rm PH}}{2r_{\rm EC}}\right)^2,
\end{equation}
where $r_{\rm PH}$ is the radius of the target photon field and $r_{\rm EC}$ is the distance between the source of the target photon field and the EC emitting region. The cartoon representation of such formalism is shown in figure \ref{fig:EC_PH_cartoon}. There is a super massive black hole (SMBH) at the centre of the system surrounded by the BLR region, a dusty torus with a relativistic jet lying perpendicular to the plane of the accretion disk. \sa{As the emitting region is far from the central zone, it can be considered as a point source when viewed from the emitting region.}

The energy dependence term in Equation \ref{eq:characteristic} i.e., the energy-loss rate of an electron due to the EC process in the presence of an external photon field in the jet comoving frame \sa{is adopted from \citep{Khangulyan_2014} and} is given by,
\begin{equation}\label{eq:f(E)_EC}
    f(\epsilon) = \frac{c_{\rm e} log\left(1 + 0.722 \times \frac{\epsilon} {c_{\rm e}}\right)}{1 + \left(c_{\rm e} \times \frac{\epsilon}{0.822}\right)},
\end{equation}
where c$_{\rm e} = 4.62$ and $\epsilon = 4\gamma\Theta$, $\gamma =\frac{E}{m_e c^2}$ and $\Theta =\frac{k_{\rm B} T_{\rm cmv}}{m_{\rm e}c^2}$ are the energy of the electron and temperature of the photon field in the jet comoving frame, respectively, given in the units of m$_{e}$c$^{2}$. Such approximation for estimation of $f(\epsilon$) is applicable in both Thomson and Klien Nishina limits and provides accuracy of an order of 1$\%$. \sa{Besides, as was already noted, the dilution factor takes care of the angular distribution of seed photon field, and the temperature of the black body inherently is dependent on the energy density of the photon field. These presumptions are addressed by the analytical expression that is presented here.}
In this work, we have solved Equation \ref{eq:characteristic} numerically by implementing the fourth order Runge-Kutta method. The particle spectra and the energy distribution in each bin are updated using the formalism provided in \cite{Vaidya_2018}. \sa{A brief discussion on the interplay among the multiple radiation mechanisms in the cooling process is discussed in Appendix \ref{sec:coolingtime}.}

\subsection{Calculation of External Compton emissivity}\label{sec:method_ICemis}

In this section, we describe the quantification of emissivity due to EC scattering because of the interaction of relativistic electrons with a given radiation field. In the present work, we have assumed that the black body photon field is mono-directional and adopted an expression of the interaction rate with relativistic electrons in the co-moving jet frame.  

For the case of mono-directional black body radiation field, the interaction rate in the co-moving frame is given by \citep[Eq. 11 of][]{Khangulyan_2014},
\begin{equation}
  \frac{dN_{\rm ph}}{d\omega_{\rm sc} dt} = \frac{2r_0^2m_e^3c^4\kappa_{\rm cmv}\Theta^2}{\pi \hbar^3 \gamma^2}\left[\frac{\sa{q}^2}{2(1-\sa{q})}F_1(x_0) + F_2(x_0)\right] \label{eq:Int_rate}
\end{equation}
where, $\omega_{\rm sc} = \sa{q}\gamma$ is the up-scattered photon energy in the units of m$_{\rm e}$c$^{2}$ and $x_0 = \sa{q}/(1-\sa{q})\epsilon_{\vartheta}$, with $\epsilon_{\vartheta}$ = 2$\gamma$\,$\Theta$(1 - $\cos\vartheta_{\rm e})$, $\cos\vartheta_{\rm e}$ = $\frac{\cos\theta_{\rm obs} - \beta}{1 - \beta\cos\theta_{\rm obs}}$ is the electron's direction and $\theta_{\rm obs}$ is the observing angle. The two analytic functions $F_1$ and $F_2$ are estimated with 1\% accuracy as demonstrated in \cite{Khangulyan_2014}.

We further assume that the relativistic electrons \sa{within each macro-particle} are isotropically distributed in the co-moving jet frame. \sa{The assumption of such isotropic electron distribution is, indeed, a simple zeroth order approximation as it allows to simply determine the electron direction and has been used by \cite{Dermer_1995, Georganopoulos_2001}.} We can therefore estimate the total emissivity of the electron population by convolving the above interaction rate with electron distribution as follows:
\begin{equation}\label{eq:EC_jnu_phy}
    \nu_{\rm sc} j^{\prime}_{EC} (\nu_{\rm sc}) = \int_{\gamma_{\rm min}(t)} ^{\gamma_{\rm max}(t)} \left[\frac{dN_{\rm ph}}{d\omega_{\rm sc} dt}\right] \omega_{\rm sc} m_{\rm e} c^{2} N(\gamma) d\gamma 
\end{equation}

In the above equation, the scattered frequency $\omega_{\rm sc}$ = $\frac{h\nu_{\rm sc}}{m_{\rm e}c^{2}}$, where $\nu_{\rm sc}$ can be obtained from the observed frequency, $\nu_{\rm sc}$ = $\frac{\nu_{\rm obs}}{D}$ with $D$ = [$\Gamma(1 - \beta\cos\theta_{\rm obs})]^{-1}$ is the Doppler boosting factor. The limits to the integration are the minimum and maximum value of the electron energy in terms of m$_{\rm e}$c$^{2}$ and are time dependent as the electron energy is evolving following the radiative loss equation~\ref{eq:characteristic}. 

The emissivity in the observer frame in the units of erg~cm$^{-3}$ s$^{-1}$ Hz$^{-1}$ str$^{-1}$ is given by,
\begin{equation}\label{eq:Jnu_Dop2}
    j^{\rm obs}_{\rm EC}(\nu_{\rm obs}) = D^2 \, j^{\prime}_{\rm EC}(\nu_{\rm sc}).
\end{equation}

Finally, we can deposit the EC emissivity obtained for each macro-particle on to the grid cells so as to give the grid distribution of $j^{\rm obs}_{\rm EC}(\nu_{\rm obs}, \bvec{r})$ and then can obtain specific intensity maps by integrating along the line of sight.

\section{Slab Jet Simulation}\label{sec:slabjet}

To understand the fundamental impact of high energy emission mechanisms such as the EC process on the energy spectra and emission features, at first, we focus on a two-dimensional relativistic slab jet. For that purpose, in section \ref{sec:SJ_setup&Emis}, we have discussed the numerical setup required to simulate a 2D relativistic slab jet along with emission modelling particulars. \sa{Section \ref{sec:sj_emission} discusses the results obtained from our simulations in the context of multi-band emission. Further, a brief discussion on evolution of particle spectra due to several radiative processes in addition to the validation of our numerical algorithm is provided in Appendix \ref{sec:sj_spectra}.}

\subsection{Model setup}\label{sec:SJ_setup&Emis}

In this work, the numerical simulations are carried out using the relativistic magneto-hydrodynamic (RMHD) module of the \texttt{PLUTO} code \citep{Mignone_2007_PLUTO}. 
\sa{The numerical setup is the same as described in \cite{Vaidya_2018}. The summary of the computational details and initial value of the parameters are given in a tabular form (see table \ref{tab:2D_setup})}. The \sa{simulation} is carried out in a Cartesian domain having a dimension of x = (0, L) and y = (-L/2, L/2).
The jet \sa{of width $l$} is under-dense
in comparison to the ambient medium \sa{with a density ratio $\eta$}. Furthermore, an initial uniform magnetic field is applied along the x-direction corresponding to a plasma beta of value 10$^{3}$. \sa{The flow has a velocity in the same direction where the ambient medium is static. The initial values of the magnetic field strength and Lorentz factor is given in table \ref{tab:2D_setup}.} 
Following \cite{Bodo_1995}, we apply a perturbation in the $y$ component of the velocity that leads to the development of Kelvin–Helmholtz instability (KHI) and consequently generates shocks at the jet boundary or at the vortices. The total computational time of the simulation is 200\,t$_{\rm sc}$, which is equivalent to a duration of 0.13\,Myr in physical units with t$_{\rm sc}$ = 6.52 $\times$ 10$^{2}$ years. Further, the unit density on the central axis of the jet and the unit length scale is chosen to be 1.661 $\times$ 10$^{-28}$ gm/cm$^3$ and L = 2000$\pi$ pc, respectively.

\begin{table}
    \centering
    \caption{\sa{Summary of the parameters of slab jet simulation.}}
    \label{tab:2D_setup}
    \setlength{\tabcolsep}{6pt} % Default value: 6pt
    \renewcommand{\arraystretch}{1.5} % Default value: 1
    \begin{tabular}{|c|c|}
    \hline \hline
    \multicolumn{2}{|c|}{Numerical details and initial conditions} \\
    \hline 
    Geometry & Cartesian  \\
    Dimension & 2D  \\
    Resolution & 384 $\times$ 384 \\
    Boundary condition & Periodic in x, outflow in y \\
    \hline
    Density ratio ($\eta$) & 10$^{-2}$  \\
    Plasma beta parameter ($\beta$) & 10$^{3}$  \\
    Magnetic field strength & 6m$G$ \\
    Lorentz factor ($\Gamma$) & 5 \\
    Jet width ($l$) & 400 pc \\
    \hline
    \end{tabular}
\end{table}

\begin{figure*}
    \centering
    \includegraphics[scale=0.26]{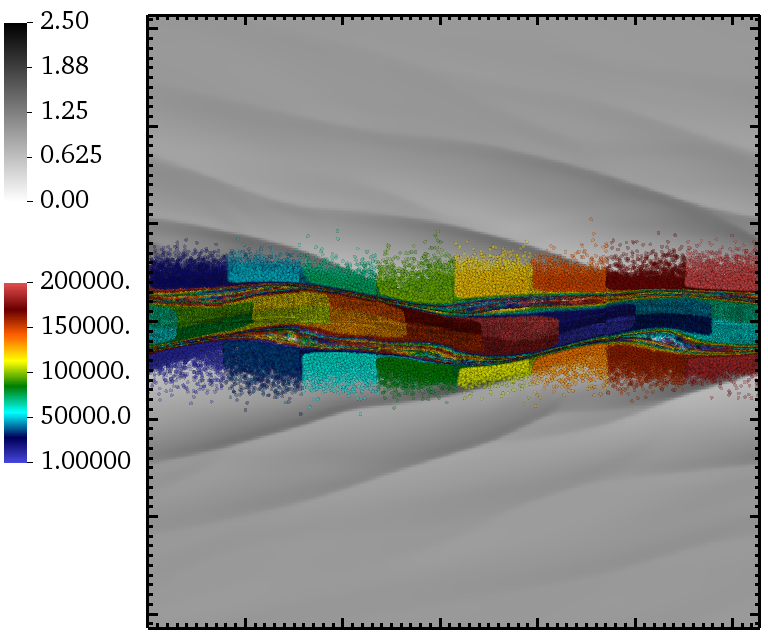}
    \includegraphics[scale=0.26]{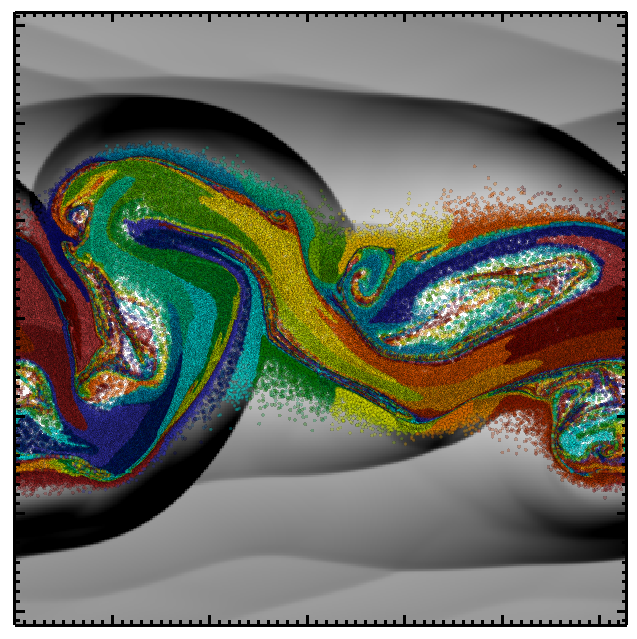}
    \caption{2D representation of particle distribution at t/t$_{\rm sc}$ = 150 (left) and 200 (right) overlaid with the fluid density (grey color). The colour bars show the fluid density and particle unique identities.}
    \label{fig:particle_sj}
\end{figure*}

We have performed two different simulations;
\begin{itemize}
    \item Case-1 : Without the inclusion of energy loss due to EC,
    \item Case-2 : Including the energy loss due to EC.
\end{itemize}
These simulations are used to present a comparative analysis among the energy loss terms given in Eq. \ref{eq:characteristic} and the corresponding results are given in section \ref{sec:sj_emission} and \ref{sec:sj_spectra}. To study the effects of different energy loss mechanisms and their impacts on the emission signatures, we introduced 2\,$\times$\,10$^5$ number of macro-particles at the initial time of the simulation. The distribution of the Lagrangian particles allows a complete sampling of the system, as shown in figure \ref{fig:particle_sj}. Here, the particles are coloured based on their unique identity, ranging from 1 to the maximum number of macro-particles. The background on which these particles are over-plotted represents the fluid density in grayscale. The left and right panels of figure \ref{fig:particle_sj} are the snapshots taken at times t/t$_{\rm sc}$ = 150 and 200, respectively. The combined evolution of Lagrangian macro-particles and the fluid density from our 2D runs clearly shows that the slab jet is sufficiently sampled during the evolution. Further, one can observe that the vortices of the KHI are responsible for mixing the particles within the slab jet as it evolves. An initial spectra with a power law index of $p$ = 6 is considered for each macro-particle, with energy cutoffs of $\gamma_{\rm min}$ = 10$^{2}$ and $\gamma_{\rm max}$ = 10$^{8}$ distributed over 256 bins. \sa{It should be noted that the choice of particle spectra is arbitrary. However, since the spectra will eventually flatten as a result of the particles being shocked, we have initially chosen a steeper index.} The synchrotron emissivity for an initial power law particle spectra is calculated using Eq. 37 of \cite{Vaidya_2018} and the EC emissivity is calculated using Eq. \ref{eq:EC_jnu_phy} of section \ref{sec:method_ICemis}. All the emissions are obtained at the time t/t$_{\rm sc}$ = 200 (particle distribution is shown in the right panel of figure \ref{fig:particle_sj}) with an observer making a 5$^{\circ}$ angle with respect to the z-axis (pointing out of the plane).

\subsection{Effect on multi-band emission maps}\label{sec:sj_emission}

\begin{figure}
    \centering
    \includegraphics[scale=0.48]{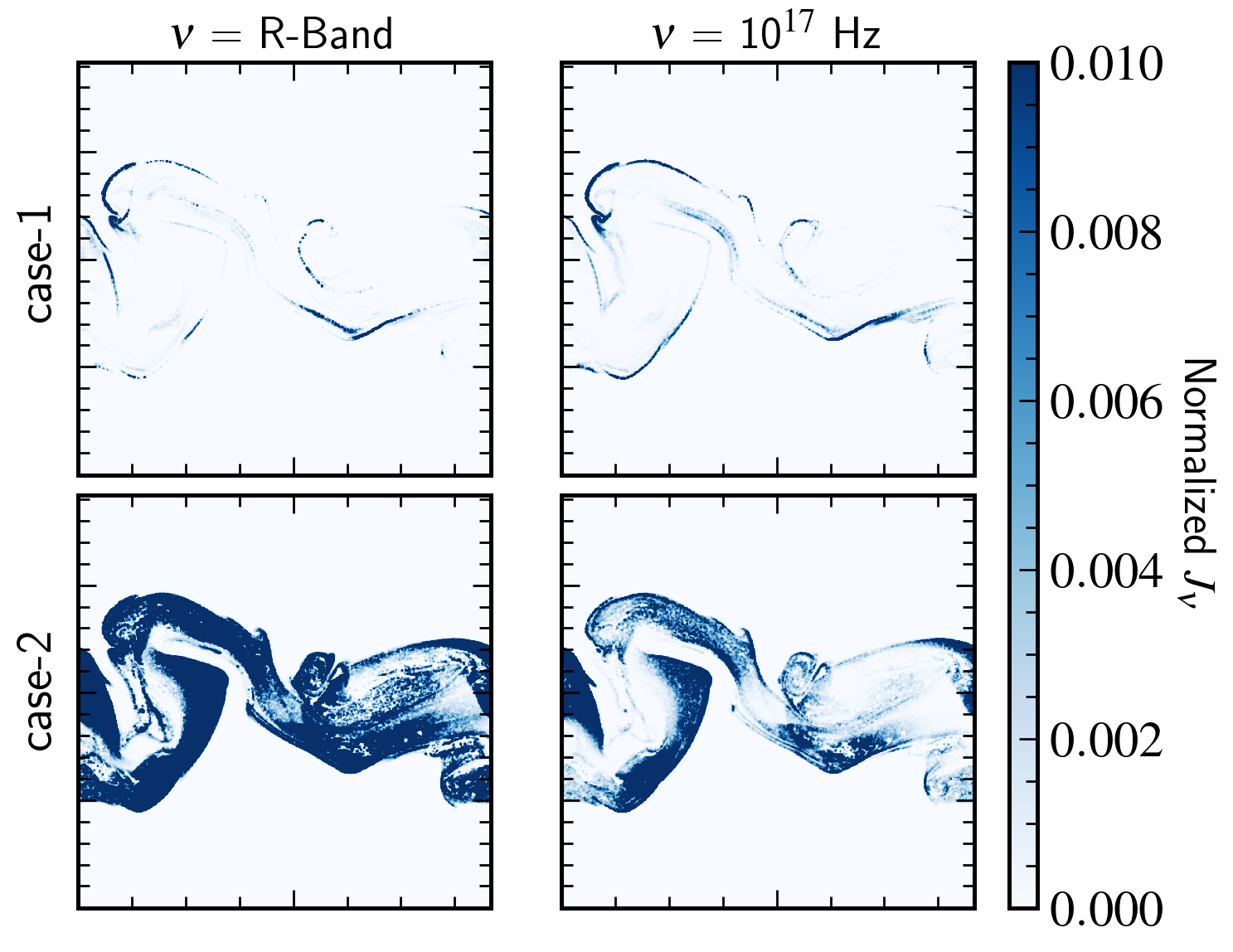}
    \caption{Emissivity slices of slab jet for case-1 and case-2 at an observing frequencies $\nu$ = R-Band and 10$^{17}$ Hz at t/t$_{\rm sc}$ = 200. Here, the emissivity values are normalized to their individual maximum values.}
    \label{fig:Jnu_compare}
\end{figure}

Another way of studying the effect of energy loss due to the EC process is to estimate the emissivity at different frequencies. The temperature of the seed photon field is assumed to be 1000 K \sa{(corresponding to typical temperature of IR photons from the torus \citep{Donea_2003_torusTemp, Malmrose_2011, Oyabu_2017_torusTemp})} for the calculation of EC emissivity, and the dilution factor value is estimated to be $\kappa$ = 10$^{-7}$ by considering the r$_{\rm PH}$ $\approx$ 0.1 pc \sa{(representative scales associated with inner structure of torus \citep{urry_1995})} and r$_{\rm EC}$ > 10 pc. The top and bottom panels of figure \ref{fig:Jnu_compare} show the emission maps for case-1 and case-2 in optical R-Band (left) and 10$^{17}$ Hz (equivalent to 0.3 keV) (right) respectively. The emissions shown at both frequencies are normalised to their individual maximum values and given in the units of ergs s$^{-1}$ cm$^{-3}$ Hz$^{-1}$ str$^{-1}$. The maximum values of emissivities for R-Band and 10$^{17}$ Hz are 1.18 $\times$ 10$^{-40}$ $\&$ 1.74 $\times$ 10$^{-47}$ (top) and 3.44 $\times$ 10$^{-41}$ $\&$ 6.25 $\times$ 10$^{-41}$ (bottom) respectively.
Such a comparative analysis of emission due to different radiative processes exhibits different structures observed in the jet.
The emission maps for case-1 and case-2 show significant differences at both frequencies (top and bottom left panels). For case-1, the synchrotron emission is mostly coming from the shocked region. However, for case-2, an extended emission is observed at both R-Band and 10$^{17}$ Hz with the inclusion of loss due to EC compared to the case with loss due to only synchrotron and IC-CMB.
This suggests that for the chosen set of values of photon field temperature and $\kappa$, the addition of loss due to the EC process significantly affects the particle spectra. \sa{It can be seen from the time evolution plot of particle spectra for a single test particle (see the bottom panel of figure \ref{fig:time_evol_Ne} of the appendix) which} in turn reflects on the emission signatures. The lower energetic particles lose energy significantly because of EC loss compared to the high energy particles because of synchrotron loss.
These particles up-scatter the lower energetic photons and are responsible for the observed extended emission. In case-1, however, there are no such electrons to emit optical and X-rays throughout the jet, hence we only observe emission at the shocked region.
Note that, for the scenarios with lower photon field temperature or lower $\kappa$ values compared to the considered ones, the effect of loss due to EC will be reduced. As a result, the EC emission will be observed at 10$^{17}$ Hz and at further higher energies.

\begin{figure*}
    \centering
    \includegraphics[scale=0.73]{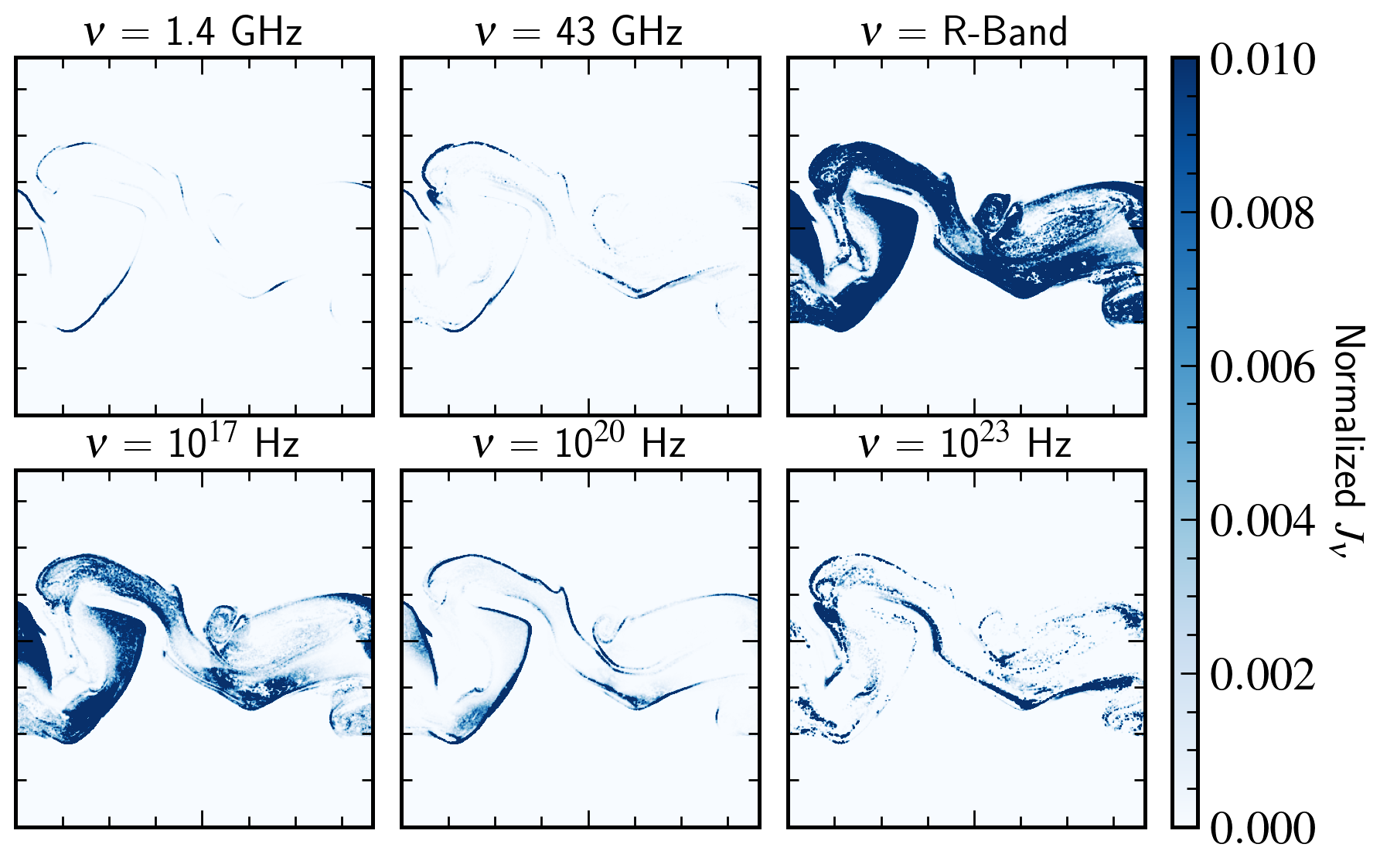}
    \caption{Emissivity slices of slab jet for case-2 for different observing frequencies at t/t$_{\rm sc}$ = 200. Here, the emissivity values are normalized to their individual maximum values.}
    \label{fig:Jnu_slices}
\end{figure*}

Additionally, we have estimated the emission at multiple frequencies, starting from radio to $\gamma$-rays. The top panel of figure \ref{fig:Jnu_slices} shows emissivity maps obtained at 1.4 GHz, 43 GHz, and optical R-Band, and in the bottom panels, the emission maps are obtained at $\nu$ = 10$^{17}$ Hz, 10$^{20}$ Hz and 10$^{23}$ Hz, respectively. The emissivity values are normalised to their individual maximum values, showing different features at different frequencies and given in the units of ergs s$^{-1}$ cm$^{-3}$ Hz$^{-1}$ str$^{-1}$. The maximum values at 1.4 GHz, 43 GHz, R-Band, 10$^{17}$ Hz, 10$^{20}$ Hz and 10$^{23}$ Hz are 1.59 $\times$ 10$^{-35}$, 5.5 $\times$ 10$^{-39}$, 3.44 $\times$ 10$^{-41}$, 6.25 $\times$ 10$^{-41}$, 1.07 $\times$ 10$^{-51}$, 1.29 $\times$ 10$^{-60}$ respectively.
As expected, in the radio bands, the synchrotron emission gets diminished with an increase in the observing frequencies, and the features get enhanced at 43 GHz due to the lower normalization value.
In the optical and 10$^{17}$ Hz (X-ray), an extended emission is observed throughout the jet in addition to the shocked emission at the sheared regions. The emission weakens due to radiative cooling and additional features are visible at higher energies.
In the EC scattering, the lower energetic photons gain energy and get up-scattered, where the frequency of the photon after up-scattering is given by $\nu_{\rm up}$ $\approx$ $\gamma^{2}\nu_{0}$ \citep{Longair2011}. In our study, the considered photon field temperature of T = 1000 K corresponds to IR photons with $\nu_{0}$ $\approx$ 10$^{13}$ Hz. According to the recipe given in \cite{Longair2011}, the IR photons would require electrons of Lorentz factor $\gamma$ $\sim$ 10$^{1}$ - 10$^{3}$ to obtain emission at R-Band and 10$^{17}$ Hz. Similarly, electrons with $\gamma$ $\sim$ 10$^{4}$ and $\gamma$ $\sim$ 10$^{5}$ - 10$^{6}$ are required to obtain emissions at 10$^{20}$ Hz and 10$^{23}$ Hz respectively. It is expected to have a higher number of lower energetic electrons since initially we considered a power-law particle spectra from $\gamma_{\rm min}$ = 10$^{2}$ to $\gamma_{\rm max}$ = 10$^{8}$. In addition to that, due to the strong effect of loss due to EC, many particles would lose energy and reaches up to $\gamma_{\rm min}$ = 10$^{1}$. Such distribution of lower energetic electrons is responsible for the extended optical and X-ray emission. As the number of particles reduces with the increasing $\gamma$ of electrons, the emission at higher frequencies also reduces.

\section{Plasma column Simulation}\label{sec:plasma_column}

Blazar jets exhibit multi-timescale variability and emit non-thermal emissions covering the entire gamut of the electromagnetic spectrum. These jets are highly magnetised and are prone to undergo several MHD instabilities during their propagation in space, and could possibly trigger jet radiation and particle acceleration. In \cite{Acharya_2021}, we investigated the physical configurations preferable for the formation of kink mode instability by performing relativistic MHD simulations of a plasma column using the \texttt{PLUTO} code. The plasma column depicts a portion of an AGN jet about a few tens away from the central engine and hence magnetically dominated and relativistic in nature.
Additionally, connecting the dynamics of the plasma column with its emission features, we found a correlated trend between the growth rate of kink mode instability with its flux variability, obtained from the simulated light curves. We would like to stress that, in the previous work, we have focused on only the optical band and have estimated the synchrotron emission by using a static particle spectra. However, in this work, we have improved our emission modelling approach by using an evolving particle spectra and including a high energy emission mechanism for a better understanding of multi-wavelength emission properties.

\subsection{Model setup}

The simulations are initialised in the computational domain of size 8$R_{\rm j}$ $\times$ 8$R_{\rm j}$ $\times$ 12$R_{\rm j}$ with $R_{\rm j}$ = 0.5 being the radius of the cylindrical plasma column.
A force-free initial magnetic field profile is chosen for the equilibrium condition \citep{Mizuno_2011, Anjiri_2014} and is given by,
\begin{equation}
B_{\rm z}\frac{dB_{\rm z}}{dR} + \frac{B_{\rm \phi}}{R}\frac{d}{dR}(RB_{\rm \phi}) = 0,\label{eq:force_free}
\end{equation}
where, R = $\sqrt{x^2 + y^2}$ is the radial position in the cylindrical coordinate system. Here, the magnetic field in the radial direction is B$_{\rm r}$ = 0 with B$_{\rm z}$ $\&$ B$_{\rm \phi}$ being the poloidal $\&$ toroidal magnetic field components. The column has a flow velocity in the z-direction, given by a bulk Lorentz factor $\Gamma_{\rm z}$ = 5, where the ambient medium is static. The initial equilibrium of the system is perturbed by a velocity provided in the radial direction. Note that the perturbation is provided in such a way that the number of kink that fit into the simulation box is $n$ = 4.
Following \cite{Acharya_2021}, the on-axis column magnetisation (magnetic-to-matter energy density) in the relativistic form is given by $\sigma_{0} = \frac{B_{0}^{2}}{\rho_{\rm c}}$.
Here, B$_{0}$ and $\rho_{\rm c}$ = 1.0 are the magnitudes of the magnetic field strength and density on the axis given in the non-dimensional units, respectively. To study the effect of magnetisation on the emission features, we chose two values of $\sigma_{0}$. The other detailed description of the initial conditions and numerical methodology is provided in \cite{Acharya_2021}.
Here, the dimensionless quantities can be scaled with appropriate physical units, relevant for the present study. For the relativistic module, the unit velocity is equal to the speed of light c = 2.998\,$\times$\, 10$^{10}$ cm/s, the unit length is chosen to be  R$_{\rm sc}$ = 0.1\,pc and the unit density is $\rho_{\rm sc}$ = 1.673\,$\times$\,10$^{-24}$ gm/cm$^{3}$. The above choices set the time and the magnetic field to be in units of t$_{\rm sc}$ = 0.32\,years and B$_{\rm sc}$ = 1.374\,$\times$\,10$^{-1}$ Gauss, respectively.

To understand the effect of the dynamics of the system on the non-thermal emission, we initialised all the runs with 3\,$\times$\,10$^{5}$ number of Lagrangian macro-particles using the hybrid framework of \texttt{PLUTO} code. This methodology takes into account the effects of microphysical processes on the distribution functions of emitting particles and, subsequently, on emissivities. To model the non-thermal emission, similar to the slab jet problem, we consider a power law particle spectra with an initial power law index $p$ = 6 with energy bounds $\gamma_{\rm min}$ = 10$^{2}$ and $\gamma_{\rm max}$ = 10$^{8}$ distributed over 256 bins with equal bin widths on logarithmic scale.
The synchrotron and EC emissivities are obtained using the equations mentioned in section \ref{sec:SJ_setup&Emis} while considering an observer making a 5$^{\circ}$ angle with respect to the axis of the column (jet). In this work, we have studied two reference cases with $\sigma_{0}$ = 10 and 1, named Ref\_s10 and Ref\_s1. In addition, for comparative analysis, different seed photon field temperatures and $\kappa$ values are considered, and the details of these simulation setups are given in table \ref{tab:run_details}.

\begin{table}
    %\centering
    \caption{All simulation run details are given column wise as runs ID, magnetisation value on the axis ($\sigma_{0}$), temperature of the seed photon field (T) and the dilution factor ($\kappa$) considered while estimating the EC emission. In all the cases, we have a decreasing pitch profile with sound speed c$_{s0}/c$ is 0.127. These values are given at the initial time of the simulation.}
    \label{tab:run_details}
    \centering
    \setlength{\tabcolsep}{12pt} % Default value: 6pt
    \renewcommand{\arraystretch}{1.5} % Default value: 1
    \begin{tabular}{|c|c|c|c|}
    \hline \hline
    Runs ID & $\sigma_{0}$ & T (in Kelvin)  & $\kappa$  \\  
    \hline
    Ref\_s10 & 10.0 & 5000 & 10$^{-2}$ \\ 
    Ref\_s10\_A & 10.0 & 2000 & 10$^{-2}$ \\ 
    Ref\_s10\_B & 10.0 & 5000 & 10$^{-3}$ \\ 
    Ref\_s1 & 1.0 & 5000 & 10$^{-2}$ \\ 
    \hline
    \end{tabular}
\end{table}  

\subsection{Results}

In this section, we have mainly focused on the results obtained from the Ref\_s10 and Ref\_s1 cases and discussed their multi-wavelength properties (simulated light curve $\&$ SED). The variability and correlation studies provide useful information regarding the location and size of different emitting regions and sources of high energy emission in differently magnetised environments.

\subsubsection{Emission maps}

\begin{figure*}
    \centering
    \includegraphics[scale=0.58]{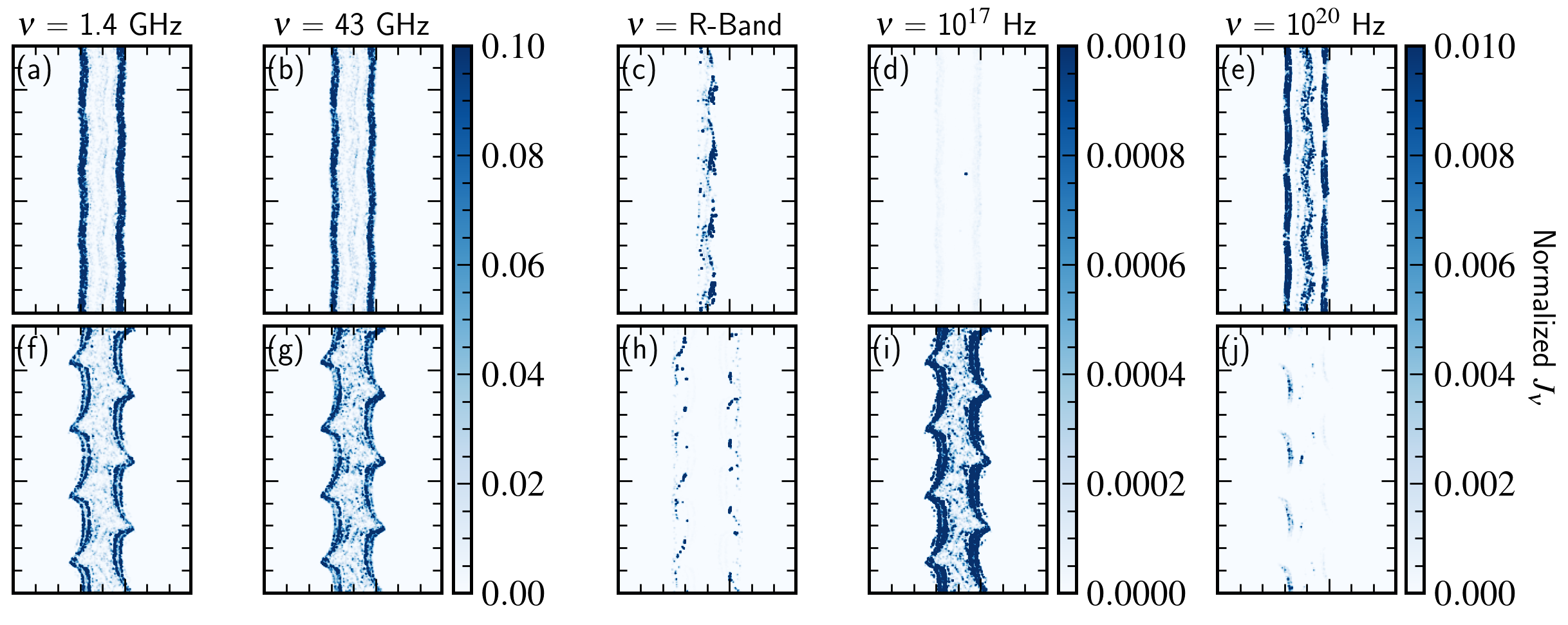}
    \caption{X-Z cuts of emissivity slices for Ref\_s10 case at observing frequencies $\nu$ = 1.4 GHz (a), 43 GHz (b), R-Band (c), 10$^{17}$ Hz (d) and 10$^{20}$ Hz (e) at time stamps t/t$_{\rm sc}$ = 26 (top) and 70 (bottom). The emissivity values are normalized to their maximum values and given in the units of erg s$^{-1}$ cm$^{-3}$ Hz$^{-1}$ str$^{-1}$. The normalized values of 3.42 $\times$ 10$^{-22}$, 4.23 $\times$ 10$^{-22}$,  3.94 $\times$ 10$^{-25}$, 8.31 $\times$ 10$^{-27}$, 3.4 $\times$ 10$^{-35}$, 2.83 $\times$ 10$^{-22}$, 1.74 $\times$ 10$^{-22}$,  4.76 $\times$ 10$^{-26}$, 6.71 $\times$ 10$^{-29}$, 3.4 $\times$ 10$^{-38}$ for the panels a, b, c, d, e, f, g, h, i and j respectively.}
    \label{fig:Jnu_Ref_t70}
\end{figure*}

The contribution of loss due to EC on the emission features has already been shown in section \ref{sec:sj_emission} in the context of a 2D relativistic slab jet. Here, in figure \ref{fig:Jnu_Ref_t70}, we have shown the X-Z cuts of normalised emissivity slices at Y = L$_{y}$/2 for Ref\_s10 case at t/t$_{\rm sc}$ = 26 (top) and 70 (bottom). At both time stamps, in radio bands (see panels a, b, f, and g), strong emission is coming from the boundary in addition to the extended emission observed throughout the column.
With the onset of the instability at t/t$_{\rm sc}$ = 26, there are formation of shocks. As a result, we observed localised synchrotron emission near the axis of the column in R-band. However, at this early evolution time of the instability, the loss rate is not sufficient to exhibit emission at 10$^{17}$ Hz. At higher energy (see panel e), EC is the dominant process and exhibits an extended emission throughout the column.
Due to the high growth rate of the instability in Ref\_s10, the emissivity maps show prominent structures of the formation of kink at t/t$_{\rm sc}$ = 70.
At this time stamp, in the optical band, the localised synchrotron emission is mostly observed at the shocked positions or at the kinked portion of the column. At 10$^{17}$ Hz, an extended EC emission is observed throughout the jet from the contribution of lower energetic particles, as a result of strong EC cooling. Additionally, there is a contribution of localised synchrotron emission from the shocked positions.

In the low magnetised environment, Ref\_s1 case, due to the trans-Alfv\'enic nature of the flow, a mixing of kink and KHI is expected. However, as a consequence of lower magnetic field strength, the growth of the instability is not enough to exhibit notable morphological changes (emission maps are not shown here). The synchrotron emission gets diminished while observing at radio frequencies due to the radiative cooling effect. In the optical band and 10$^{17}$ Hz, the very few localised shocked particles emit synchrotron emission in addition to the broad or extended EC emission throughout the column. A comparably lower emission is observed at 10$^{20}$ Hz as a result of EC cooling.

\subsubsection{Multi-wavelength light curve}\label{sec:Multi_LC}

\begin{figure*}
    \centering
    \includegraphics[scale=0.5]{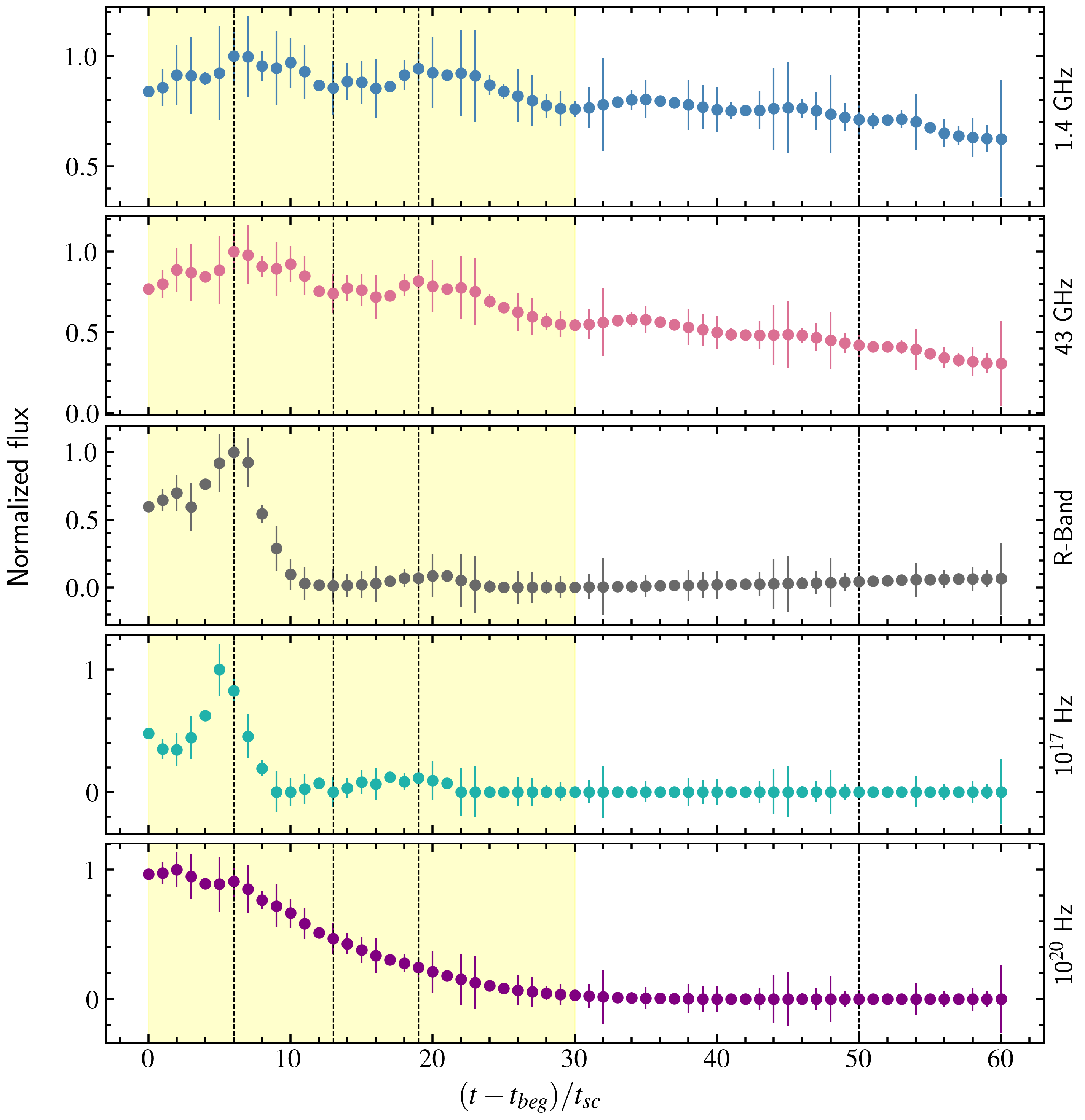}
    \caption{Simulated light curve for the Ref\_s10 case for an observer making 5$^\circ$ angle with the axis of the column. The observing frequencies are $\nu$ = 1.4 GHz, 43 GHz, 1.4 $\times$ 10$^{14}$ Hz (R-Band), 10$^{17}$ Hz and 10$^{20}$ Hz respectively from top to bottom. The vertical black dotted lines and the shaded regions correspond to different activity state of the system (discussed in section \ref{sec:Multi_LC}). The flux values from top to bottom are normalized to their maximum values 2.64 $\times$ 10$^{-17}$, 2.52 $\times$ 10$^{-17}$, 2.66 $\times$ 10$^{-21}$, 5.13 $\times$ 10$^{-24}$, 1.54 $\times$ 10$^{-31}$ respectively and given in the units of ergs s$^{-1}$ cm$^{-2}$ Hz$^{-1}$.}
    \label{fig:LC_Ref_5deg}
\end{figure*}

One of the ways to understand the emission characteristics is to estimate the total integrated flux and study its time varying properties. All relativistic effects, such as relativistic boosting and light travel effects, have been taken into account while estimating the integrated flux. \sa{In particular, the emissivities are obtained in the observer frame by multiplying the comoving emissivity by the square of the Doopler boosting factor (for example see Eq. \ref{eq:Jnu_Dop2}). Furthermore, we have used a realistic slow-light approximation \citep{Bronzwaer_2018} to collect information about the light as it travels through the grid cells of the simulation box.} We have shown the simulated multi-wavelength light curve for the Ref\_s10 case in figure \ref{fig:LC_Ref_5deg}, where the fluxes are normalised to their maximum values and provided in the units of erg s$^{-1}$ cm$^{-2}$ Hz$^{-1}$. The light curve is shown from t/t$_{\rm sc}$ = 20 to 80, corresponding to a period of $\approx$\,20 years in physical units. During this period, the instability is in its evolving state, and hence, it is favourable to study the associated emission signatures during this phase. The yellow highlighted or shaded region indicates a period when the light curve shows distinctive features and mainly corresponds to the linear growth phase of the instability. Later on, during the non-linear growth of the instability, the magnetic field structure becomes chaotic and a decaying phase of the light curve is observed. The black vertical dotted lines are plotted to indicate different activity states of the system and are labelled as state-1, state-2, state-3, and state-4. In state-1, at t/t$_{\rm sc}$ = 26, an enhancement of the flux can be seen in the given frequencies shown in figure \ref{fig:LC_Ref_5deg}. In state-2, at t/t$_{\rm sc}$ = 33, a less active state is observed in all energy bands compared to state-1. In state-3, at t/t$_{\rm sc}$ = 39, a semi-harmonized behaviour along with flux enhancement is observed, suggesting a moderately active state of the system. Lastly, in state-4, at t/t$_{\rm sc}$ = 70, a comparably low activity state is observed since not much variation in flux can be noticed at that particular time. The spectral behaviour of the system at these particular times is also given in section \ref{sec:sed}. In radio bands, a moderately variable emission is seen during the yellow shaded duration. However, in the optical R-Band and 10$^{17}$ Hz, a characteristic transient feature is seen at t/t$_{\rm sc}$ = 26 followed by a sharp decay in the light curve. Further, at t/t$_{\rm sc}$ = 39, few particles encounter shocks again leading to the high flux state in R-Band and 10$^{17}$ Hz compared to the state-2 (t/t$_{\rm sc}$ = 33) where the previously shocked/emitting particles were in a cooling phase. Such transient characteristics are due to the contribution of localised synchrotron emission as a result of the generation of shocks. Note that the flux values are very low at the higher energies compared to the lower energies and the certain extreme jump in flux in R-Band and 10$^{17}$ Hz is due to the few number of particles encountering localised shocks.
A decaying nature of the light curve is observed at 10$^{21}$ Hz ($\gamma$-ray).

\begin{figure*}
    \centering
    \includegraphics[scale=0.5]{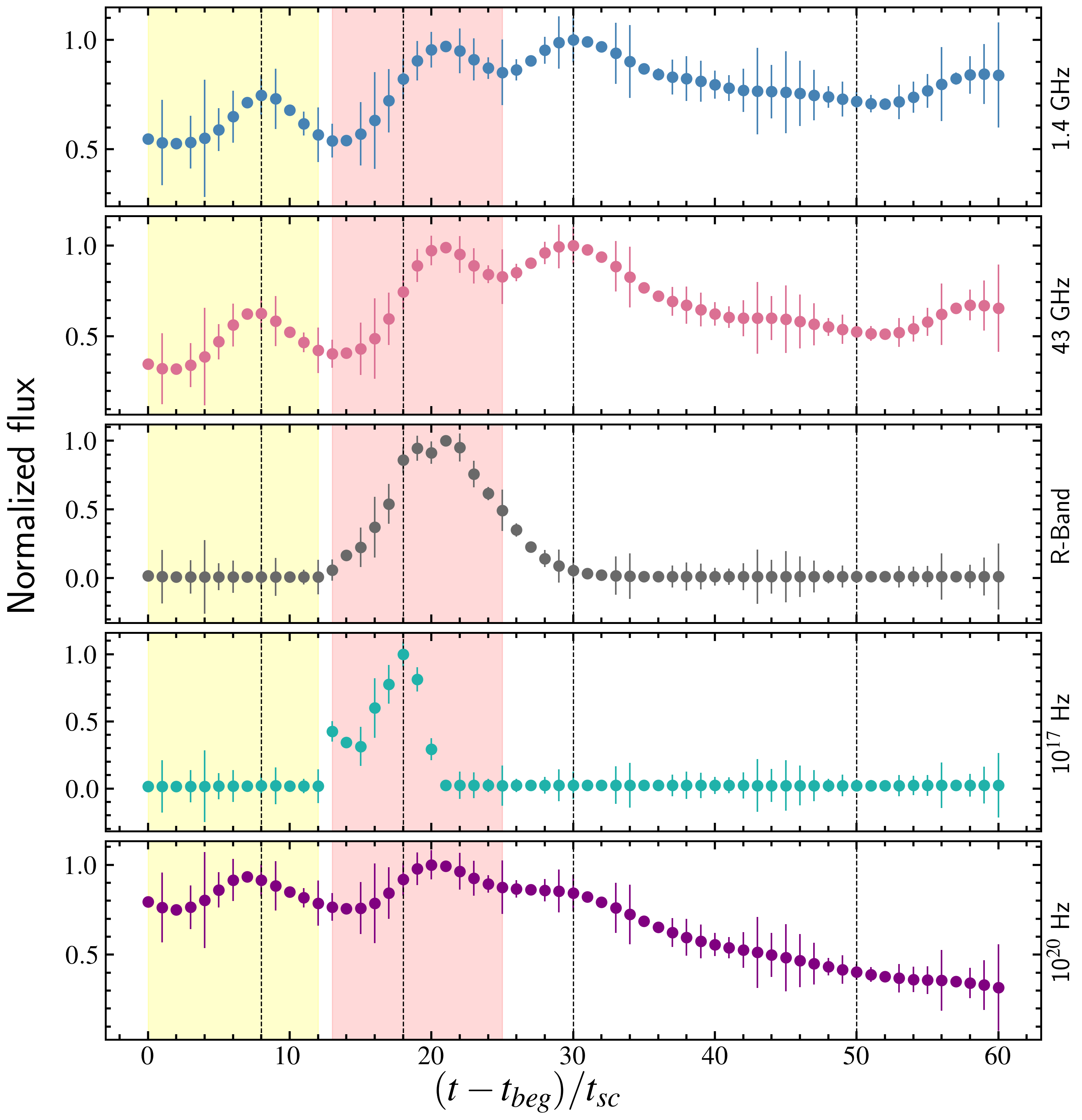}
    \caption{Simulated light curve for the Ref\_s1 case. The observing frequencies and other details are same as figure \ref{fig:LC_Ref_5deg}. The flux values from top to bottom are normalized to their maximum values  2.58 $\times$ 10$^{-18}$, 7.99 $\times$ 10$^{-19}$, 6.65 $\times$ 10$^{-23}$, 1.9 $\times$ 10$^{-25}$, 3.99 $\times$ 10$^{-32}$, respectively and given in the units of ergs s$^{-1}$ cm$^{-2}$ Hz$^{-1}$.}
    \label{fig:LC_Ref_s1_5deg}
\end{figure*}

Similarly, we have shown the simulated light curve for the lower magnetisation case, Ref\_s1, in figure \ref{fig:LC_Ref_s1_5deg}. Here also, the light curve is shown for a period that is equivalent to $\approx$\,20 years in physical units, and the flux values are provided in the units of ergs s$^{-1}$ cm$^{-2}$ Hz$^{-1}$. In this case, the growth of the instability is not sufficient to show significant structural formation.
The shaded regions and the vertical lines indicate different activity states shown by the system, and the spectral behaviour at the corresponding times is given in section \ref{sec:sed}. A moderately variable emission is noticed in all energy bands except for the R-band and 10$^{17}$ Hz emission. Similar to the Ref\_s10 case, a transient feature is observed at R-band and 10$^{17}$ Hz due to the localised synchrotron emission (see the red shaded region). However, the energy of the particles responsible for the transient optical emission is less compared to the energy of the particles responsible for the transient X-ray emission. As a result, the optical light curve decays slowly, whereas the X-ray light curve decays rapidly as a consequence of the faster radiative cooling effect.

\subsubsection{Correlation $\&$ variability analysis}\label{sec:dcf&variability}

\begin{figure*}
    \centering
    \includegraphics[scale=0.135]{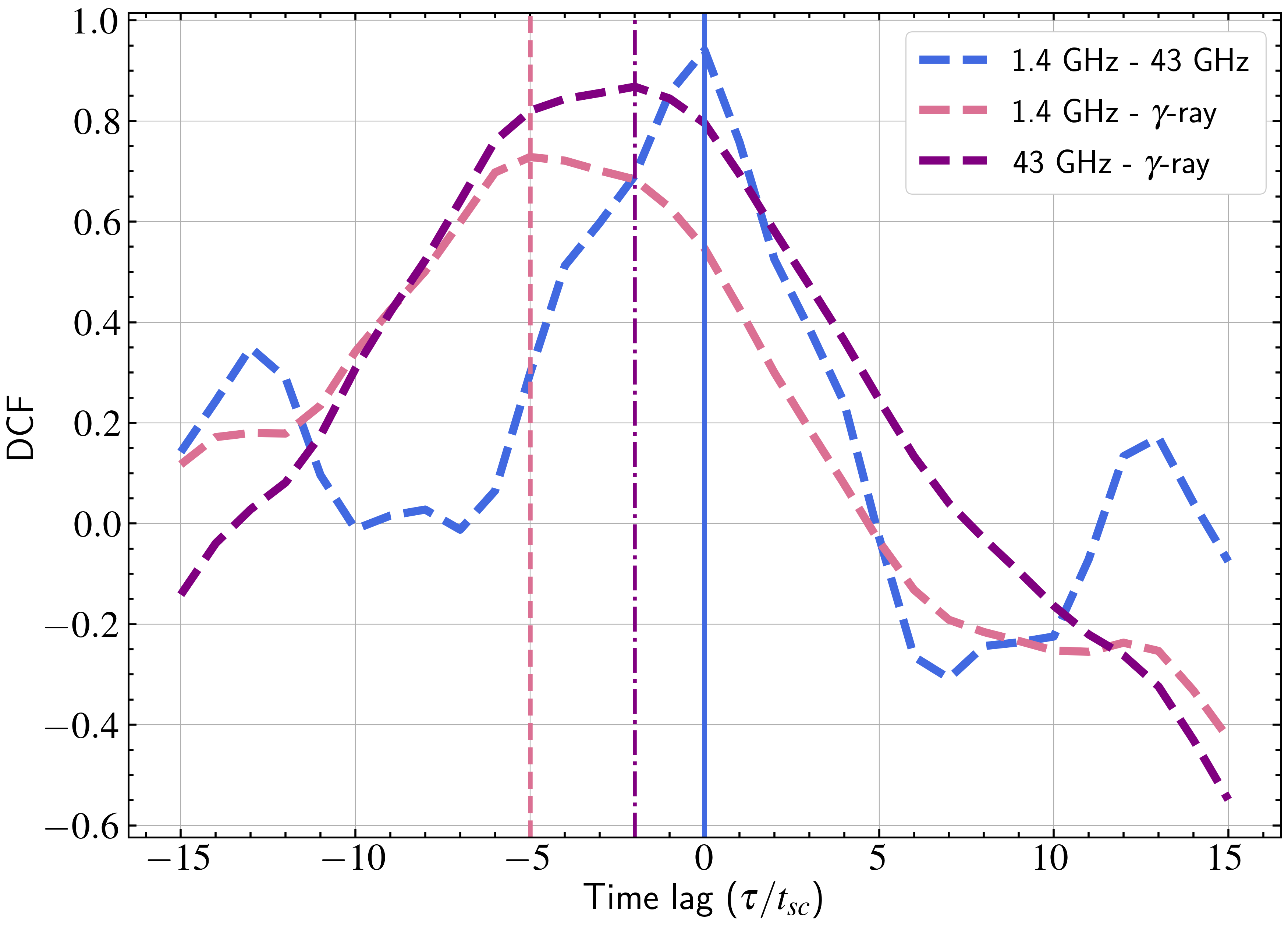}
    \includegraphics[scale=0.135]{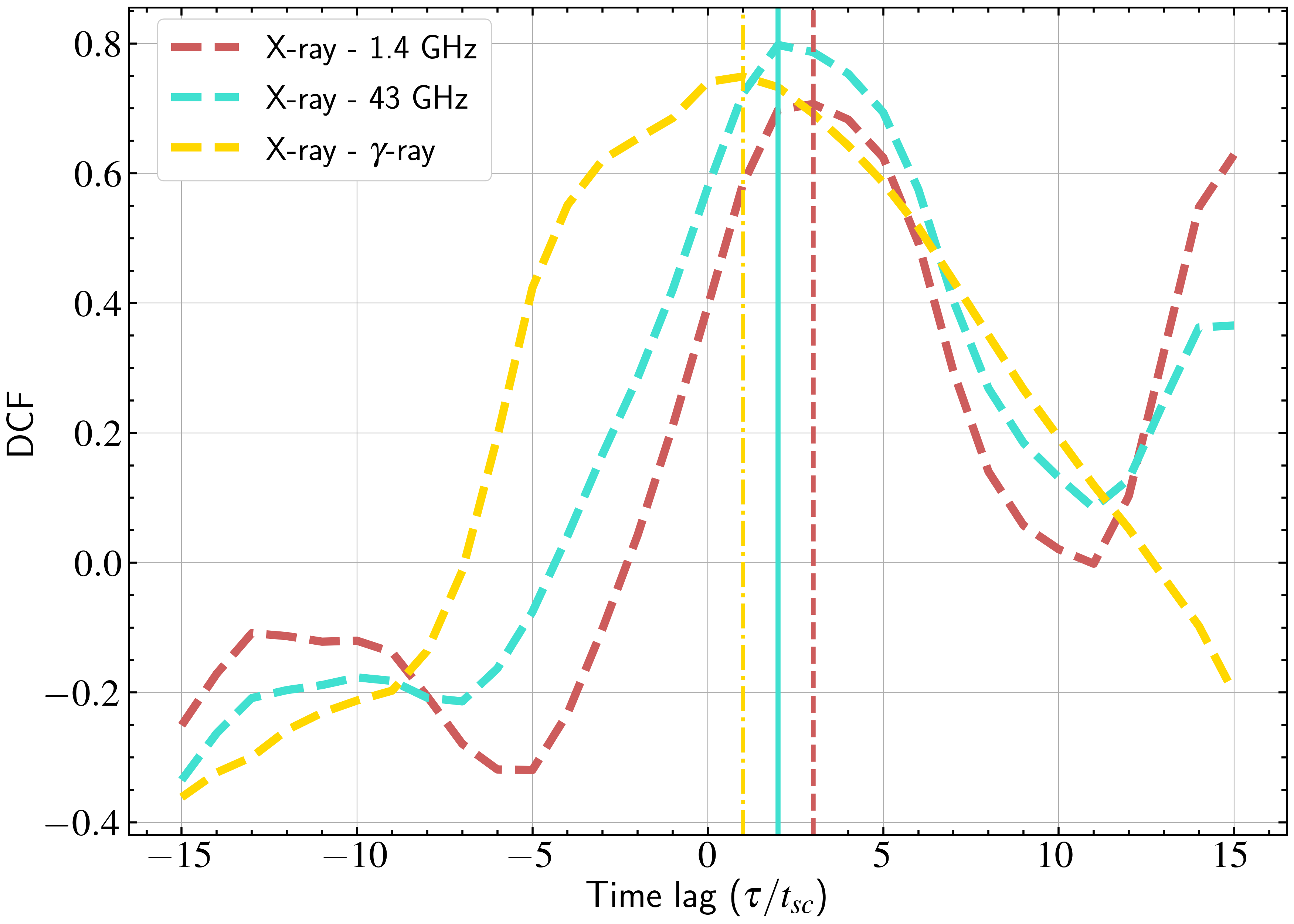}
    \includegraphics[scale=0.135]{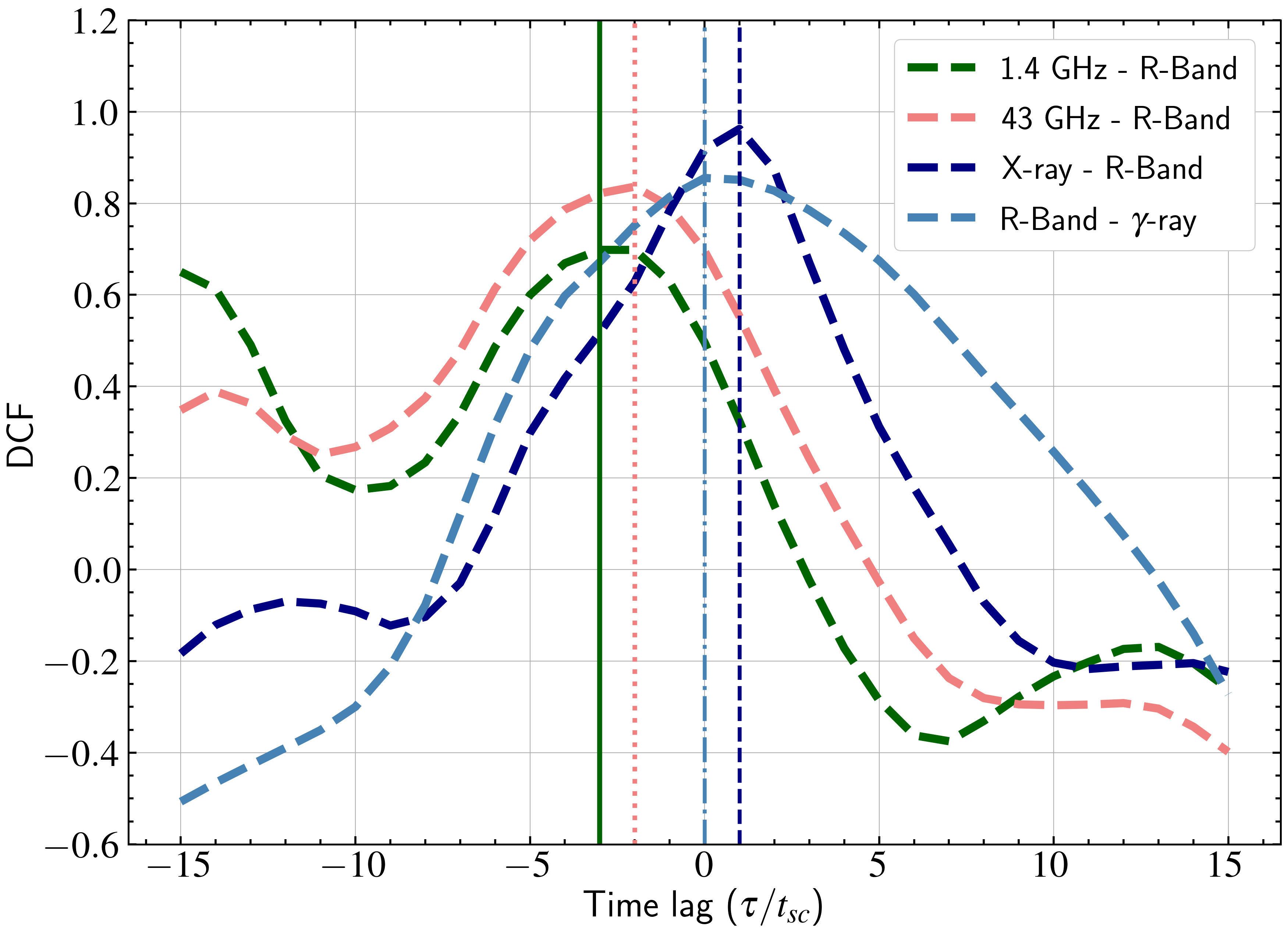}
    \caption{DCF plots between different energy bands for Ref\_s10 cases for a time duration t/t$_{\rm sc}$ = 20 to 50 (yellow shaded region in figure \ref{fig:LC_Ref_5deg}). The vertical lines indicate the lag time when the DCF of each energy band pair peaks.}
    \label{fig:DCF}
\end{figure*}

For correlation studies between these light curves, we have performed DCF (Discrete Correlation Function) \citep{edelson_krolik_dcf} for different activity states. DCF is a method for measuring correlation functions without interpolating the data in the temporal domain.
The unbinned DCF function is defined as:
\begin{equation}
UDCF_{\rm ij} = \frac{(a_{\rm i}-\bar a)(b_{\rm j}-\bar b)}{\sqrt{\left((\sigma_{a}^{2}-e_{a}^{2})-(\sigma_{b}^{2}-e_{b}^{2})\right)}},
\end{equation}
where $\bar a$ and $\bar b$ are the mean of the two discrete data sets a$_{\rm i}$ and b$_{\rm j}$, respectively. $\sigma_{\rm a/b}$ and e$_{\rm a/b}$ are the standard deviation and error associated with each data set. Note that, for simplicity, we have not added the simulated errors to the dataset i.e., e$_{\rm a/b}$ is taken as 0 while estimating the UDCF$_{\rm ij}$ values.
Further, the DCF($\tau$) can be measured by binning the result in time. Averaging over the M pairs for which $\tau - (\delta\tau)/2 < \delta t_{\rm ij} \ll \tau - (\delta\tau)/2$ , the DCF is given as:
\begin{equation}
DCF(\tau) = \frac{1}{M} UDCF_{\rm ij}
\end{equation}
Figure \ref{fig:DCF} shows the DCF plots among radio, optical R-Band, X-ray (10$^{17}$ Hz) and $\gamma$-ray (10$^{20}$ Hz) for the yellow shaded region of the Ref\_s10 case, as shown in figure \ref{fig:LC_Ref_5deg}. We found that the emissions at radio frequencies (1.4 GHz $\&$ 43 GHz) show a strong correlation with zero time lag, where each data corresponds to $\sim$\,4 months of binned data. This suggests that the emitting regions for the radio bands are co-spatial within that period \sa{\citep{Baliyan_2001}}. Additionally, during this period, the $\gamma$-ray emission is correlated with radio emission with 5t$_{\rm sc}$ and 2t$_{\rm sc}$ lag for 1.4 GHz and 43 GHz, respectively (left panel of figure \ref{fig:DCF}). \sa{Earlier study by \cite{Max_Moerbeck_2014} also found such lag in radio emission from the $\gamma$-ray emission.}
The DCF plots shown in the middle panel of figure \ref{fig:DCF}, indicate a correlation between the X-ray emission with radio band emissions with a lag of 3t$_{\rm sc}$ and 2t$_{\rm sc}$ for 1.4 GHz and 43 GHz, respectively. This suggests that the emission in the X-ray band comes prior to the emission in the radio bands, i.e., the highly energetic shocked electrons emit first in the X-ray followed by the radio emission. Similarly, there is a correlation between X-ray and $\gamma$-ray with a positive lag of 1t$_{\rm sc}$. Finally, the DCF plots of R-Band with all other frequencies are shown in the right panel of figure \ref{fig:DCF}. The plots indicate the correlation of R-Band emission with radio, X-ray, and $\gamma$-ray emission with different time lags.
X-ray and R-Band emissions have a 1t$_{\rm sc}$ lag, whereas optical emission is 3t$_{\rm sc}$ and 2t$_{\rm sc}$ ahead of 1.4 GHz and 43 GHz, respectively. The optical and $\gamma$-ray emission are strongly correlated with zero lag. Note, during this period of t/t$_{\rm sc}$ = 20 to 50, rise in the X-ray band emission is followed by the emission at other energy bands.
The lags observed between frequencies suggest that the emitting regions are at least c$\Delta \tau$ apart within the $\sim$\,4 months period.

\begin{figure*}
    \centering
    \includegraphics[scale=0.135]{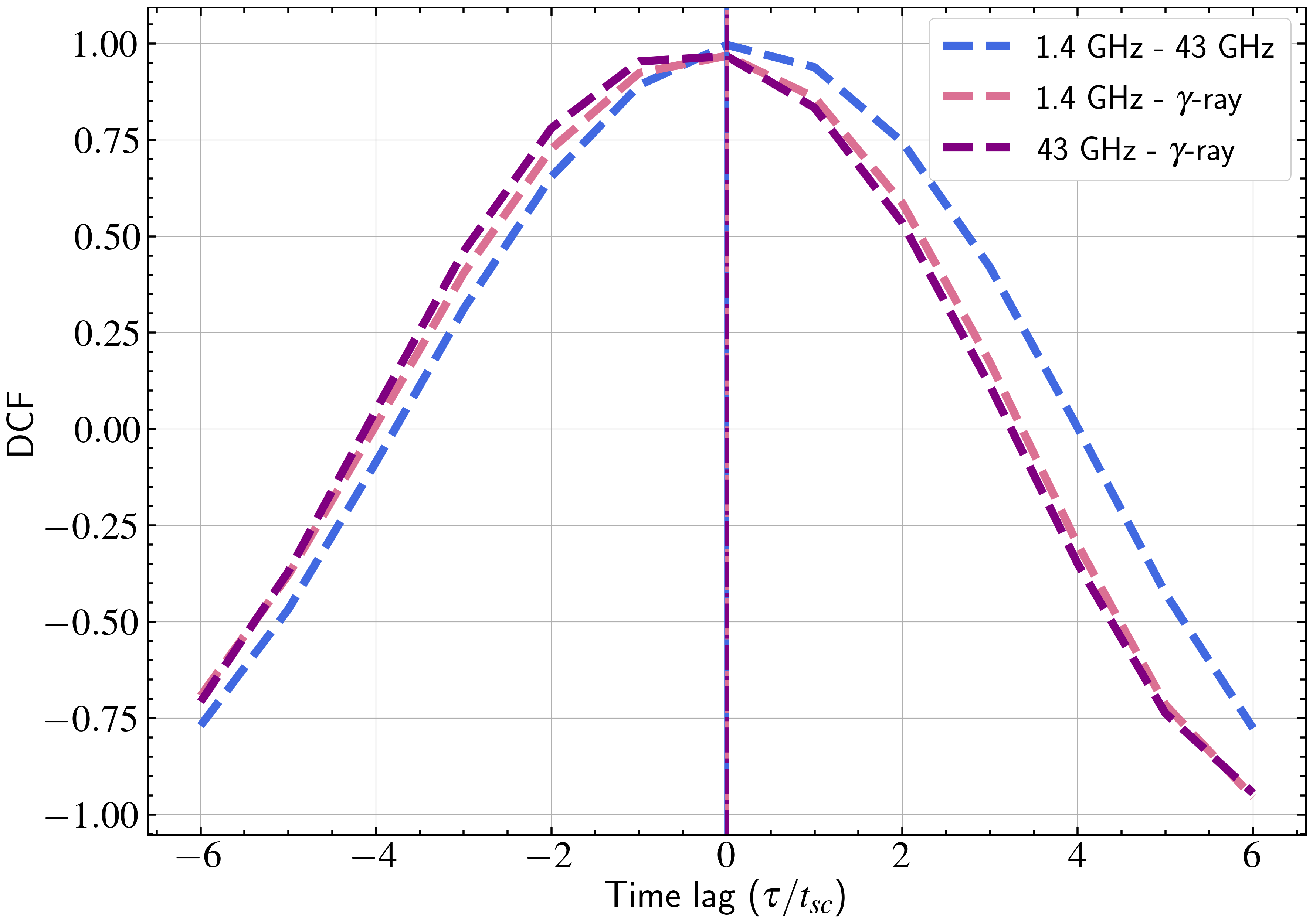}
    \includegraphics[scale=0.135]{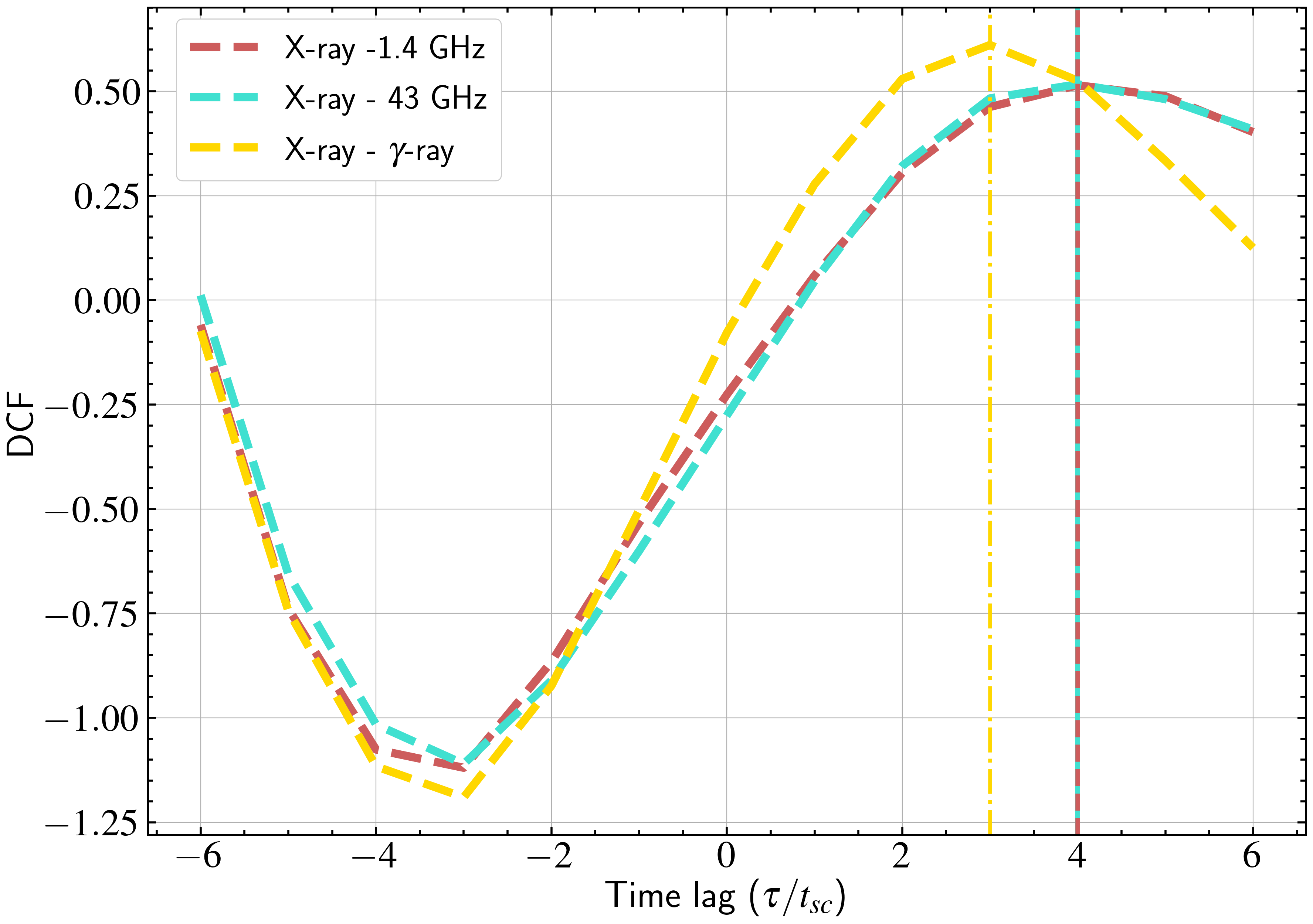}
    \includegraphics[scale=0.135]{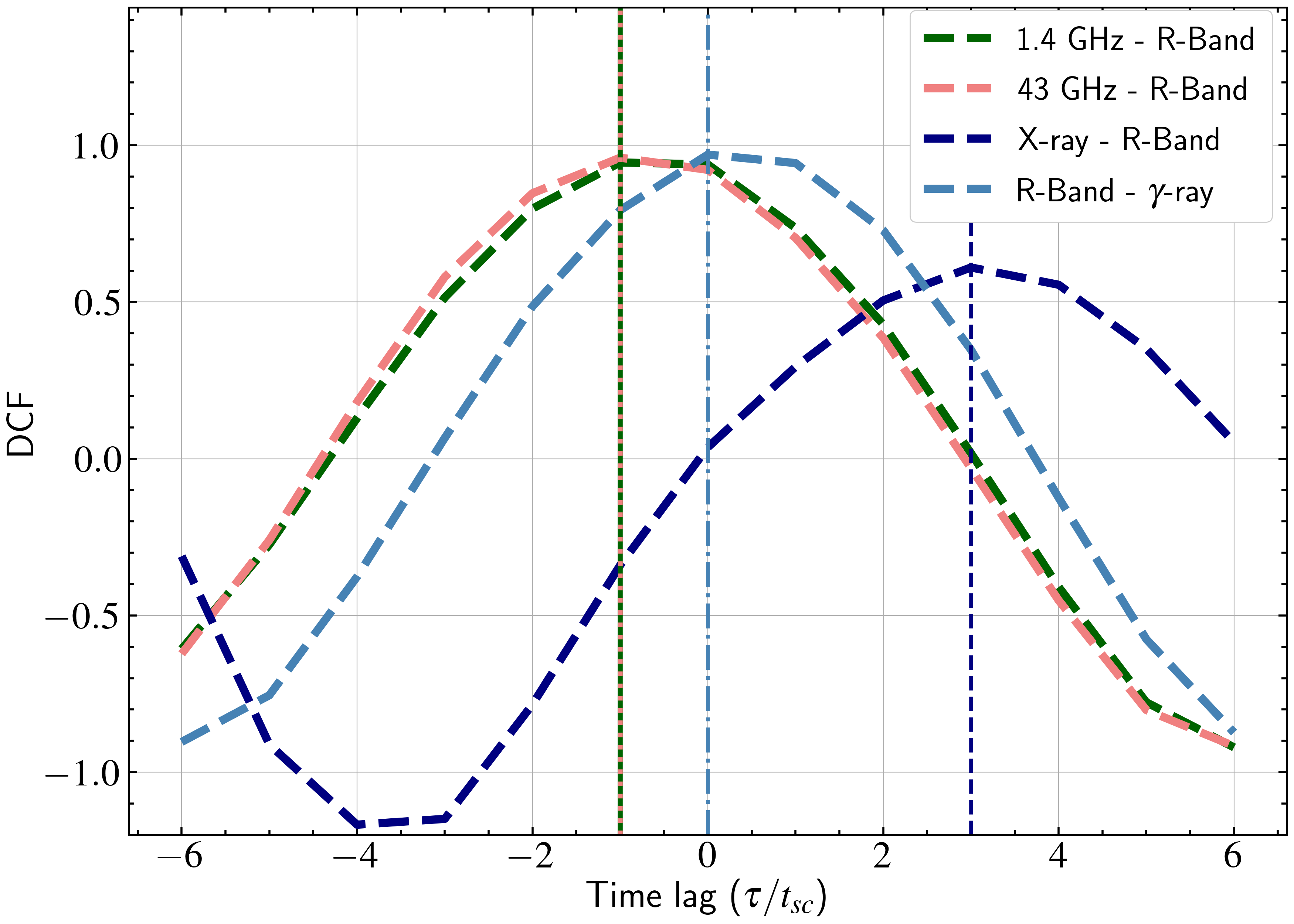}
    \caption{DCF plots between different energy bands for Ref\_s1 cases for a time duration t/t$_{\rm sc}$ = 33 to 45 (red shaded region in figure \ref{fig:LC_Ref_s1_5deg}). The vertical lines indicate the lag time when the DCF of each energy band pair peaks.}
    \label{fig:DCF_Ref_s1}
\end{figure*}

Similarly, in figure \ref{fig:DCF_Ref_s1}, the DCF plots are shown for the Ref\_s1 case for the red shaded region of figure \ref{fig:LC_Ref_s1_5deg}.
During the X-ray transient phenomena, a strong correlation with zero lag is seen among the radio bands with the $\gamma$-ray emission. This suggests that within this duration of t/t$_{\rm sc}$ = 33 to 45, the radio and $\gamma$-ray emitting regions are co-spatial with each other. \sa{Similar correlated emission between radio and $\gamma$-ray was previously observed by \cite{Ramakrishnan_2015}.} During the same period, the $\gamma$-ray emission is also strongly correlated with R-Band emissions with zero lag. But the optical emission is 1t/t$_{\rm sc}$ ahead of radio emissions due to contribution of both synchrotron and EC. In addition, X-ray emission is leading the radio and $\gamma$-ray emissions with a positive lag of 4t/t$_{\rm sc}$ and 3t/t$_{\rm sc}$, respectively. Further, X-ray emission leads the R-Band emission with a time lag of 3t/t$_{\rm sc}$ during the X-ray transient activity state.
This suggests that the recently shocked particles create another population of highly energetic electrons that emit in X-rays due to the synchrotron process. Afterwards, due to radiative cooling it loose energy and emits in low energy bands. Then, the radiated out lower energetic population gives the emission in the $\gamma$-ray band due to IC scattering.

To quantify the variability in the light curves, we estimated the relative variability amplitude (RVA) or the variability index \citep{Kovalev_2005, KSingh_2019}, defined as:
\begin{equation}
RVA = \frac{F_{\rm max} - F_{\rm min}}{F_{\rm max} + F_{\rm min}},
\end{equation}
and the uncertainty on RVA is given by:
\begin{equation}
\Delta RVA = \frac{2}{(F_{\rm max}+F_{\rm min})^{2}}\sqrt{(F_{\rm max}\Delta F_{\rm min})^{2} + (F_{\rm min}\Delta F_{\rm max})^{2}},
\end{equation}
where $F_{\rm max}$ and $F_{\rm min}$ are the maximum and minimum values of the simulated flux with $\Delta F_{\rm max}$ and $\Delta F_{\rm min}$ uncertainties, respectively. For the purpose of statistical analysis, we have also simulated the error bars for each simulated flux value. \sa{It is important to note that these are not the result of simulations or numerical uncertainties. Refer to \cite{Acharya_2021} for a further explanation of how these random errors were generated.} The estimated RVA value for Ref\_s10 and Ref\_s1 cases for a time duration of t/t$_{\rm sc}$ = 20 to 80 at different energy bands is given below in the tabular form \ref{tab:RVA_results}.

\begin{table*}
    \centering
    \caption{Standard deviation (s.d) and relative variability amplitude (RVA) values for Ref\_s10 and Ref\_s1 cases for an observer making an angle of 5$^{\circ}$ with respect to the axis of the plasma column.}
    \label{tab:RVA_results}
    \setlength{\tabcolsep}{6pt} % Default value: 6pt
    \renewcommand{\arraystretch}{1.5} % Default value: 1
    \begin{tabular}{|c|c|c|c|c|c|c|c|c|c|c|c|}
    \hline \hline
    \multirow{2}{*}{Runs ID} &
    \multicolumn{2}{c|}{1.4 GHz} & \multicolumn{2}{c|}{43 GHz} & \multicolumn{2}{c|}{R-Band} & \multicolumn{2}{c|}{10$^{17}$ Hz} & \multicolumn{2}{c|}{10$^{20}$ Hz} \\
    %\hline
    \cline{2-11}
     & s.d. & RVA & s.d. & RVA & s.d. & RVA & s.d. & RVA & s.d. & RVA \\
    \hline 
    Ref\_s10 & 0.15 & 0.23 $\pm$ 0.16 & 0.61 & 0.53 $\pm$ 0.18 & 1.39 $\times$ 10$^{3}$ & 0.99 $\pm$ 0.15 & 6.44 & 0.94 $\pm$ 0.11 & -- & 0.99 $\pm$ 0.16 \\
    
    Ref\_s1 & 0.25 & 0.31 $\pm$ 0.18 & 0.58 & 0.51 $\pm$ 0.2 & 3.84 $\times$ 10$^{2}$ & 0.99 $\pm$ 0.18 & 0.96 & 0.71 $\pm$ 0.12 & 0.66 & 0.51 $\pm$ 0.17 \\
    \hline
    \end{tabular}
\end{table*}  

In Ref\_s10 case, at radio bands (1.4 GHz and 43 GHz), the variability increases with an increase in the observing frequency. Less variability at lower frequencies can be attributed to the continuous emission coming from the whole plasma column since the emission gets diminished at higher frequencies due to radiative cooling. At higher energy bands, typically in optical and X-ray, localised high synchrotron emission is coming from the shocked/kinked region, resulting in a very high value of RVA. In $\gamma$-ray, a gradually decaying nature is observed in the light curve, resulting in a high RVA value. In the lower magnetised case (Ref\_s1), at the lower frequencies, the variability amplitude increases. However, in optical and X-ray, a transient phenomenon is observed due to the generation of localised shocks, giving rise to high RVA = 0.99 $\pm$ 0.18 and 0.71 $\pm$ 0.12 respectively. In the $\gamma$-ray band, the variability is similar to the radio bands owing to a RVA value of 0.51.

Further, from the typical minimum variability time-scales \citep{Burbidge_1974}, the size of the emitting region in the radio bands is equivalent to the size of the plasma column as the radio emission is coming from throughout the column. However, the minimum variability time-scales obtained for optical and X-ray band is nearly 0.75t$_{\rm sc}$ in the Ref\_s10 case. This corresponds to the emitting region that covers nearly 30 number of grid cells for both optical and X-ray emission out of (160\,$\times$\,160\,$\times$\,240) number of grid cells in the whole computational domain. Similarly, in Ref\_s1 case, the size of the emitting region decreases as we progress from 1.4GHz to 43GHz. However, the optical and X-ray emissions mostly come from the shocked particles that correspond to the emitting region that encloses approximately 14 and 60 grid cells, respectively. This indicates that such shocked emissions are very localised and mainly come from the kinked or sheared regions.

\subsubsection{Spectral energy distribution (SED)}\label{sec:sed}

\begin{figure*}
    \centering
    \includegraphics[scale=0.4]{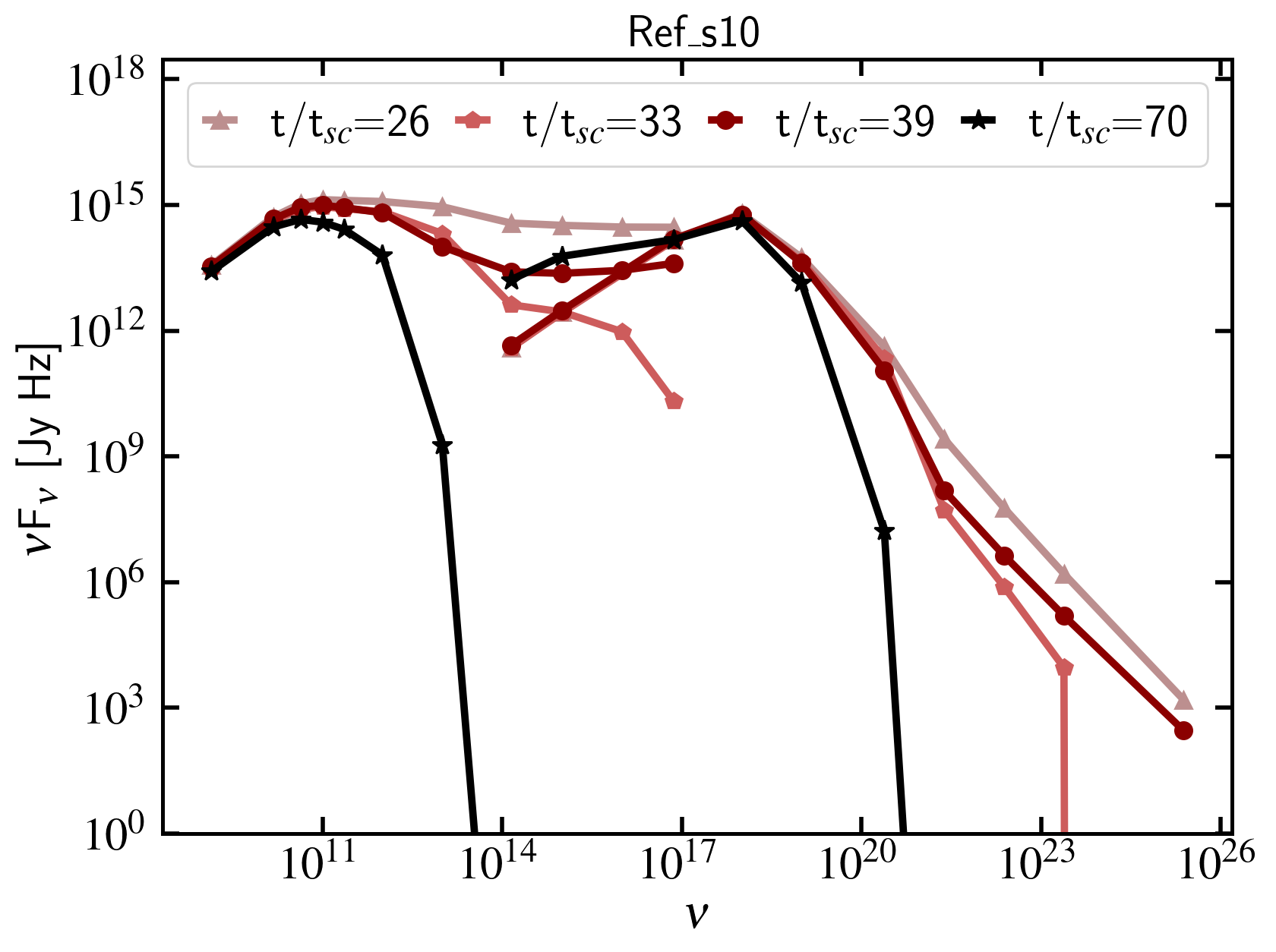}
    \includegraphics[scale=0.4]{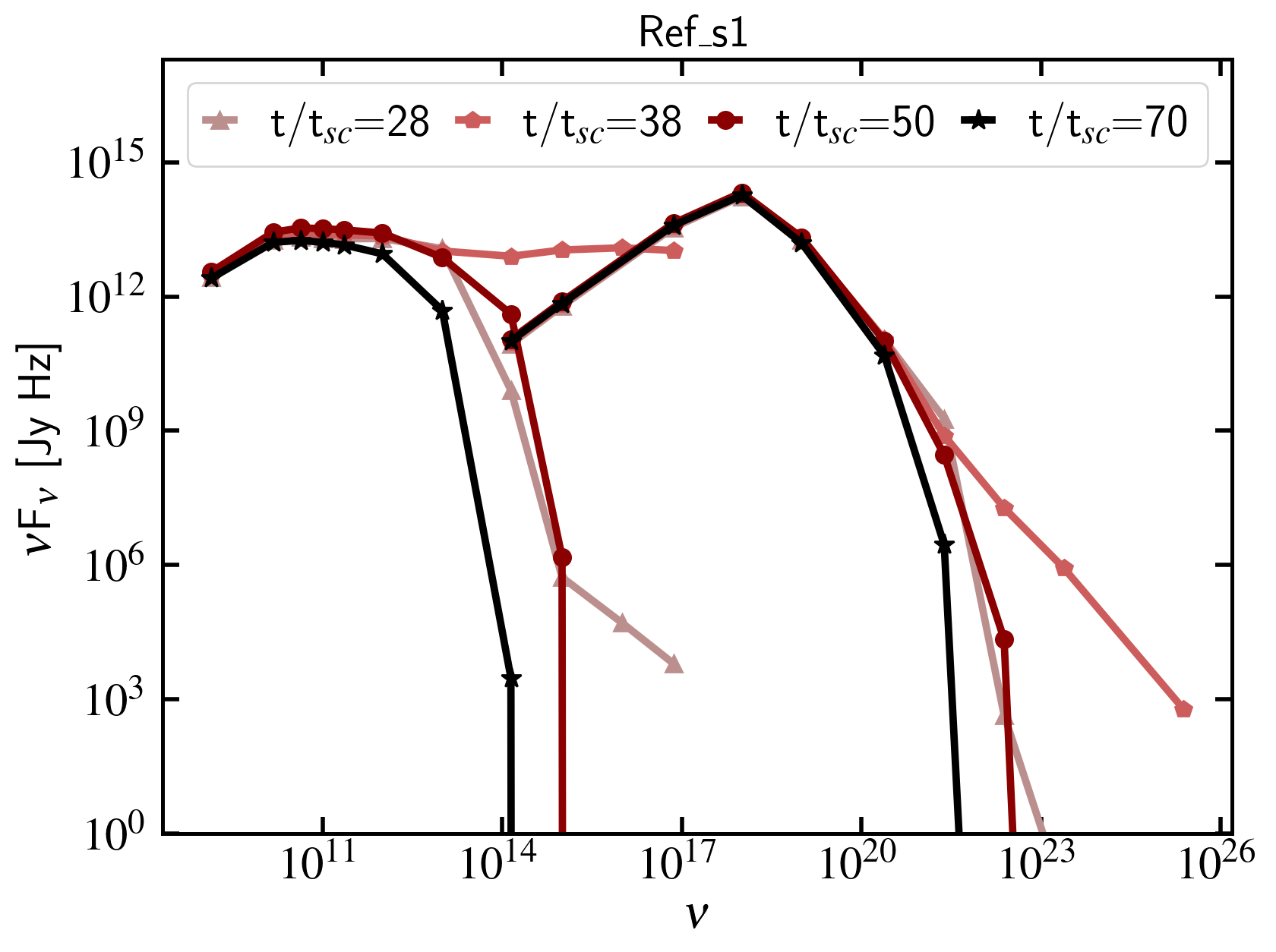}
    \caption{Left: Simulated SED of Ref\_s10 at t/t$_{\rm sc}$ = 26, 33, 39 and 70. Right: Simulated SED of Ref\_s1 cases at t/t$_{\rm sc}$ = 28, 38, 50 and 70.}
    \label{fig:SED}
\end{figure*}

We modelled the multi-wavelength SEDs of the different simulation runs by using the time-dependent fluid and spectral information. Figure \ref{fig:SED} provides the time-evolving spectral behaviour for the cases Ref\_s10 (left) and Ref\_s1 (right). The SEDs exhibit two distinct humps at each time, with the low energy hump caused by the synchrotron and the high energy hump caused by the EC process.

For the Ref\_s10 case, at the initial time of the instability (i.e., t/t$_{\rm sc}$ = 26), the synchrotron spectra is flat and the EC spectra shows a gradual decrease in the slope further 10$^{18}$ Hz.
At this time stamp, an orderly behaviour of the magnetic field lines is expected owing to the less growth of perturbation. Therefore, for an observer making a 5$^{\circ}$ angle with respect to the axis of the column, the toroidal component of the magnetic field would be dominant to provide strong synchrotron emission. In addition, the onset of instability leads to the development of shocks, generating highly energetic particles. In addition to the dynamics of the plasma column, such a mechanism can be attributed to the flattening of SED spectra, which results in a broadening of the spread of the individual humps until higher energies. At t/t$_{\rm sc}$ = 33, the shocked particles have cooled down, resulting in marginally lesser emissions at higher frequencies. However, at t/t$_{\rm sc}$ = 39, few particles again experienced shocks, and the light curve shows a comparatively low amplitude rise. Due to this, the spectra becomes flat again at this time stamp. With the evolution of instability, the magnetic energy dissipates and the total synchrotron flux values at higher frequencies decrease. As a result, the spectra become steep due to the loss of energy at the higher frequencies. At all time stamps, the high energy component of the SED behaves in a similar fashion to the low energy component as it has the same electron population responsible for both.

In the right panel of figure \ref{fig:SED}, the multi-wavelength SED for Ref\_s1 case is shown for its different activity states, as indicated in figure \ref{fig:LC_Ref_s1_5deg}.
At t/t$_{\rm sc}$ = 28, a moderately active behaviour is observed among the radio bands correlated with $\gamma$-ray emission, followed by a transient characteristic feature in the X-ray band at t/t$_{\rm sc}$ = 38 (see figure \ref{fig:LC_Ref_s1_5deg}). At this time stamp, the optical emission also shows a rise in the light curve. Such transient features in optical and X-ray bands are an indication of the generation of localised shocks. As the particles pass through shocks, the non-thermal electrons get accelerated to higher energies depending on the strength of the shocks and generate another sub-population of electrons with high electron's energy. Consequently, the particle spectra and the SED get flattened. Further, due to the high cooling rate, such shocked highly energetic particles cool faster, causing the sharp cut-off in the SED at higher energies. Like in the high magnetisation case, at all time stamps, the spectral behaviour of both the low and high energy components of the SED is similar.

\subsubsection{Parametrical comparison of SEDs $\&$ Compton Dominance}\label{sec:sed_compare}

Here, we have shown a parametrical comparison of SEDs and discussed the effects of different parameters such as \sa{$\sigma_{0}$}, T and $\kappa$ on the broad-band spectra. For this purpose, along with \sa{$\sigma_{0}$} = 10 and 1, we chose T = 5000 K and 2000 K, $\kappa$ = 10$^{-2}$ and 10$^{-3}$. \sa{The detail of each runs is given in table \ref{tab:run_details}.}

\begin{figure}
    \centering
    \includegraphics[scale=0.4]{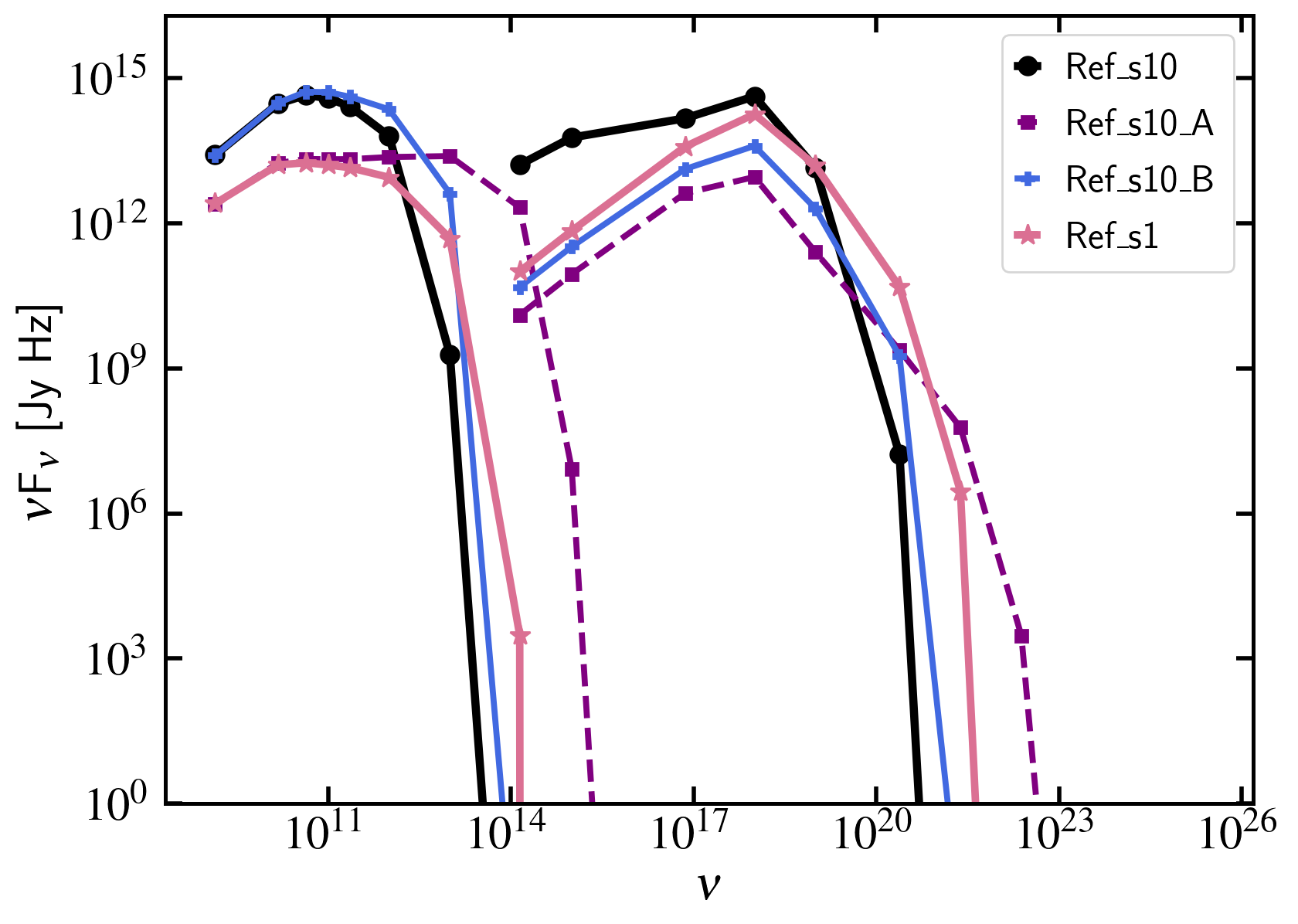}
    \caption{Parametrical comparison of simulated SEDs for an observer making 5$^{\circ}$ angle with respect to the axis of the column estimated at t/t$_{\rm sc}$ = 70.}
    \label{fig:sed_compare}
\end{figure}

In this plasma column setup, the case with a high magnetisation value (\sa{$\sigma_{0}$} = 10) is more prone to kink instability, whereas in the lower magnetisation (\sa{$\sigma_{0}$} = 1) case, a throttled growth of kink instability is observed due to the presence of weak magnetic field strength. This results in a comparatively small amplitude of the synchrotron peak. Further, inadequate dissipation of magnetic energy and less loss due to synchrotron, prompts in an orderly nature of magnetic field lines. As a result, a steady emission is observed at the lower radio frequencies up to 10$^{13}$ Hz with a broad photon spectra and then decreases. Such comparisonal behaviour of SED for $\sigma$ = 10 and 1 is shown as black (circle marker) and pink (star marker) lines, respectively, in figure \ref{fig:sed_compare}.
The Compton dominance \sa{(CD)} parameter would be one of the intriguing tools that may govern the magnetisation of the system. \sa{Typically, it is defined as the ratio of the peak of the Compton to the synchrotron peak luminosities \citep{Finke_2013, Nalewajko_2017}. In our work, we calculated the CD in two ways. Firstly, we define CD$^{\rm peak}$ as the ratio of the Compton peak (high energy hump) to the synchrotron peak (low energy hump) flux densities, noting that the ratio would stay unchanged if we calculate it from the Compton peak to the synchrotron peak luminosities. Further, we define CD$^{\rm power}$ as the ratio of EC power to the synchrotron power derived by integrating the flux values across the frequency range. It mainly gives the estimation of how much more EC emission there is than the synchrotron emission. The CD$^{\rm peak}$ and CD$^{\rm power}$ values for all the simulation runs are provided in table \ref{tab:CD}.}
For the \sa{$\sigma_{0}$} = 10 case, the \sa{CD$^{\rm peak}$} is found to be nearly 1, whereas for \sa{$\sigma_{0}$} = 1 scenario, the \sa{CD$^{\rm peak}$} is estimated to be nearly 10. Such dependencies of Compton dominance on the magnetisation parameter have previously been studied by \cite{Janiak_2015} and our results are in agreement with their outcomes. \sa{Additionally, the CD$^{\rm power}$ is higher for the lower magnetisation case since the EC component of the spectra becomes broader.}
With change in the temperature of the photon field or $\kappa$ values, the energy loss rate of the emitting particles changes. Such effect of photon field temperature and $\kappa$ values is lucid from the cooling timescale (refer figure \ref{fig:cool_time}) and the time-evolving particle spectra (refer figure \ref{fig:time_evol_Ne}). As we decrease the temperature of the photon field, the loss due to EC also decreases. As a result, it shows a significant effect on the low energy hump as well as the high energy hump of the SED. Due to less EC loss, a more number of highly energetic particles will be present to emit synchrotron emission up to 10$^{15}$ Hz. With the considered fluid and spectral information, due to the act of different radiative losses, the synchrotron emission at higher frequencies get enhanced compared to the higher photon field temperature case. Consequently, the EC emission at the higher end of the spectrum also increases. \sa{This results in broadening of the individual component of the spectra. Consequently, even though the CD$^{\rm peak}$ is found to be $\approx$ 0.4, the value of CD$^{\rm power}$ suggests that the EC power is nearly 2.5 times the synchrotron power.} The SED for this low temperature case (Ref\_s10\_A) is shown as purple dotted (square marker) lines.
The blue (plus marker) line represents the SED for the case with a decreased value of $\kappa$ (Ref\_s10\_B case) i.e., if the distance between the location of the target seed photons and the emitting region increases. Decreasing the $\kappa$ value affects the loss rate of the emitting particles and hence alters the characteristics of both humps. Like in the lower temperature scenario, the synchrotron and EC emission is more at the higher energies compared to the high $\kappa$ values, resulting in a broad photon spectra. The \sa{CD$^{\rm peak}$} for the Ref\_s10\_B case is estimated to be 0.08 \sa{and the CD$^{\rm power}$ value is obtained to be 1.2}.

\begin{table}
    %\centering
    \caption{\sa{CD$^{\rm peak}$ and CD$^{\rm power}$ values for all the simulation runs with different $\sigma_{0}$, T and $\kappa$.}}
    \label{tab:CD}
    \centering
    \setlength{\tabcolsep}{7pt} % Default value: 6pt
    \renewcommand{\arraystretch}{1.5} % Default value: 1
    \begin{tabular}{|c|c|c|c|}
    \hline \hline
    Runs ID & CD$^{\rm peak}$ & CD$^{\rm power}$ & Remarks  \\  
    \hline
    Ref\_s10 & 1.0 & 32 & Reference case \\
    \hline
    Ref\_s10\_A & 0.4 & 2.5 &  \\ 
    Ref\_s10\_B & 0.08 & 1.2 & Comparative cases  \\ 
    Ref\_s1 & 10.0 & 122 &  \\ 
    \hline
    \end{tabular}
\end{table}  

\sa{In summary, it appears that the values obtained from CD$^{\rm power}$ and CD$^{\rm peak}$ have a similar trend. However, the estimation of CD$^{\rm power}$ does not only consider the peak flux density value but also the shape of the individual spectra that is driven by various radiative cooling processes. As a result, in all cases, the values of CD$^{\rm power}$ are estimated to be greater than 1, implying higher EC power than synchrotron power.}

\section{Discussions}\label{sec:discussion}

The primary focus of this work is to investigate the impact of instability driven shocks on the long-term nature of parsec scale jets in the relativistic regime and their consequences on the multi-wavelength spectra. Previously, \cite{Acharya_2021} performed simulations of a helical jet model in the context of long-term variability and studied the dependence of the viewing angle on the observed emission using static particle spectra. In this work, the results pertaining to emission are obtained using time-dependent particle spectra, where the spectral information along with the fluid information evolves with time.
It is also important to mention that at parsec scale length, for highly magnetised and relativistic jets, IC scattering due to external photon fields (EC) is substantially more responsible for the observed emission compared to IC-CMB. On that account, in this work, we have implemented the EC process in the \sa{hybrid framework of the} \texttt{PLUTO} code.
In addition, our emission modelling approach resembles the multi-zone modelling method, where each grid cell acts as a single emitting region with time evolving fluid and spectral information.

\subsection{Incorporation of EC process}

Demonstrating the effect of the EC on the emission signatures of relativistic jets at parsec scale lengths is one of the main objectives of this paper. Note that, at this length scale, the effect of IC-CMB emission is significantly less compared to other high energy emission mechanisms such as external Compton, synchrotron self-Compton, etc. The parameterisations and approximations used in this work allow for a detailed description of the EC (IC scattering of mono-directional photons) with $\sim$\,1$\%$ precision. The executed formalism to obtain emissivities is valid in both the Thomson and Klein–Nishina limits. Our approach includes a Planckian distribution of the target photon field as suggested by \cite{Khangulyan_2014} in contrast to $\delta$-function and/or step function approximations for the seed photon fields \citep{Dermer_1992, Dermer_1993_EC, Petruk_2009, Finke_2016}.
In both the Thomson and Klein-Nishina regimes, in the presence of both isotropic and anisotropic radiation fields, the EC mechanism has been rigorously modeled by many authors that include numerically intense calculation \citep{Dermer_1992, Bottcher_1997, Hutter_2011, Hunger_2016}.
The analytical calculations performed in \cite{Zdziarski_2013} provide a high accuracy level of 0.3$\%$ for estimating the interaction rate; however, the usage of special functions in the algorithm inhibits it in its effectiveness and practical usage.
First, we characterised the energy losses due to several radiative processes through cooling time-scales (see figure \ref{fig:cool_time}). The highly energetic electrons lose energy faster and are mainly responsible for the synchrotron emission. Whereas, for EC, in the Thomson limit ($\approx$ $\gamma$ $\Theta$ $\ll$ 1), electrons lose energy like synchrotron process and in the Klein–Nishina limit ($\approx$ $\gamma$ $\Theta$ $\gg$ 1) the loss mechanism follows a logarithmic profile. \sa{Further, we have performed 2D toy model simulations that are focused on validating our numerical algorithm and understanding the effects of various parameters on the particle spectra (see figure \ref{fig:time_evol_Ne}) and multi-band emissions (see figure \ref{fig:Jnu_slices}).}

\subsection{Implications of shocks on the emission signatures}

\begin{itemize}

    \item To study the multi-wavelength nature of relativistic jets, we simulate a cylindrical plasma column with two different magnetisation values and perturb it with a kink mode instability. The higher magnetisation value is more susceptible to the kink mode instability, while in the lower magnetisation case, throttled growth of kink instability with a mixing of KHI is observed due to the presence of shear and the trans-Alfv\'enic nature of the flow \citep{Baty_2002, Acharya_2021, Acharya_2022}. With the evolution of instability, there are formation of shocks that generate another sub-population of highly energetic electrons. Especially, the shocks are formed at the kinked region in the magnetically dominated scenario, and in the lower $\sigma$ case, shear-driven shocks are produced. Such highly energetic electrons are responsible for the observed abrupt rise seen in the multi-wavelength light curves for both cases. A similar remarkable rise in the X-ray light curve is previously seen in FSRQ PKS 1222+216 \citep{Anshu_Chatterjee_2021}. Very short lived X-ray flares have also been observed in PKS 2005-489 \citep{Zhu_2018} where the variability time-scale is found to be $<$ 30s. Further, \cite{Markowitz_2022_Xray_flare} have also found modest X-ray flares in Mkn 421 that last for less than a day timescale. The kinematic analysis of high-resolution VLBI images of AGN jets reveals the presence of several blobs that are typically interpreted as recollimation shocks. A combination of multi-frequency observations with 43 GHz VLBI observations suggests that the interaction between travelling shock waves and recollimation shocks could also trigger high energy $\gamma$-ray flares \citep{Agudo_2012, Schinzel_2012}. Numerical simulations of a conical jet have shown that the interaction between recollimation shock and travelling shock may produce flaring events in both the single-dish light curves and in the VLBI observations \citep{Fromm_2016}.
    
    \item One of the important implications of shocks in our study is the appearance of X-ray orphan flares. In both high (Ref\_s10) and low (Ref\_s1) magnetised cases, a fugacious activity is observed in the X-ray band due to the generation of a sub-population of highly energetic shocked particles. \sa{Similar isolated X-ray flares have also been seen before by \cite{Abdo_2010_polarization} for 3C 273 source due to the presence of localised shocks.} During such a transient phase, the X-ray emission comes from very localised regions that lead the emissions from other energy bands. The possible explanation for such a lagged and correlated emission is that the highly energetic shocked particles lose energy due to the radiative cooling mechanisms and emit in the lower energy bands in the presence of tangled magnetic fields via the synchrotron process. These lower energetic particles get up-scattered and emit $\gamma$-rays via the EC process. \sa{In discussing the multi-wavelength correlation study for a large sample of blazars, \cite{Liodakis_2018} also offered an analogous explanation. Further, multi-wavelength correlation studies have been carried out by several authors to understand various emission mechanisms and the geometry of emitting locations. For example,\cite{Chatterjee_2012, Hovatta_2014, Fuhrmann_2014, Max_Moerbeck_2014, Cohen_2014, Ramakrishnan_2015}, and many others have conducted primarily radio, optical, and $\gamma$-ray correlation studies with the goal of constraining and locating the size of $\gamma$-ray emitting zones. Their finding supports our correlation results in the context of lag and non-lag multi-frequency emission with the time scale of lagged emission obtained from our simulated light curve lies between the observed range from several studies.}

    \item Time-evolution modelling of SED provides a better approach to study different acceleration mechanisms and constrain the physics of the jet more consistently. In our study, we found that during the transient phase observed in the flux variation, the shocked particles emit in the high frequencies and the broad-band spectra get flattened. The flattening of the spectral slope in SED is due to the re-acceleration of electrons, produced as a consequence of localised shocks. Such spectral signatures have also been observed before and studied by \cite{Micono_1999, Borse_2020}. In addition, such spectral broadening sometimes results in a shift in the peak of the synchrotron component. Recently, the multi-wavelength study of BL Lacertae shows that during the high flux state in X-rays, synchrotron emission is the dominant mechanism for X-ray production, considering the emitting region is very close to the black hole. Following the X-ray flaring activity, the X-ray component of the SED lies in the higher energy range and occurs due to the EC mechanism [Agarwal et. al. (in prep)]. Similar results were also found before for BL Lacertae by \cite{Bottcher_2003_BLLacertae}.
    
\end{itemize}

Particle acceleration due to magnetic reconnection is more effective in studying fast variability scenarios in the case of blazar jets \citep{Narayan_2012_JetinJet, Giannios_2013, Shukla_2020}. However, the current work does not specifically focus on the fast variability nature \citep{Aharonian_2007, Albert_2007, Ghisellini_2008_FastTeV, Giannios_2009, Barkov_2012, Pryal_2015}. Typically, the origin of intra-day/intra-night variable signatures in $\gamma$-ray are attributed to magnetic reconnection. Previous studies have demonstrated the role of kink in generating current sheets that could initiate fast magnetic reconnection, providing an efficient way of particle acceleration in AGN and gamma-ray-burst jets \citep{Chandra_Singh_2016, Bodo_2021, Kadowaki_2021, Dalpino_2021}. However, it is also important to highlight the effects of shocks generated at the shear interface due to the onset of instability on the emission signatures.

\subsection{Note on Compton dominance}

Magnetic fields threading the central rotating objects are predominantly responsible for the formation of relativistic jets. These jets are converted from initially Poynting flux dominated flows to matter dominated flows during their propagation in space \citep{Sikora_2005}. MHD instabilities may play a promising role in this process. In this work, we have focused on two different scenarios: one with dominant kink mode instability and the other one with mixing of both the kink mode and shear-driven KH instability. Both instances differ from each other in magnetisation values. The jet magnetisation parameter is not only important in understanding the dynamical evolution of jets but also crucial in governing the dominant particle acceleration mechanism. The Compton dominance parameter can be a measure of the magnetisation of the jet \citep{Janiak_2015}. \cite{Sikora_2005} studied the dependence of jet magnetisation on the Compton dominance as a function of the geometry of the emitting region and location of the seed photon source. They found that the Compton dominance is higher for a lower magnetisation value, irrespective of the geometry and distance of the photon source from the blazar emitting region. Similar results on Compton dominance and jet magnetisation were also suggested by \cite{Nalewajko_2017}. In our work, from the presented broad-band spectral analysis, we found Compton dominance\sa{, obtained from the peak values of flux densities (CD$^{\rm peak}$),} of approximately 10 for $\sigma$ = 1 case, whereas for $\sigma$ = 10 case, it is found to be nearly 1. Such a value of Compton dominance is also observed in FSRQs in contrast to BL Lacs due to the presence of an external photon field. In addition to different magnetisation values, we performed simulations with different photon field temperatures and $\kappa$ values.
From our analysis, we found that with the decrease in the temperature of the photon field, the \sa{CD$^{\rm peak}$} decreases and the spread of the spectrum for both humps increases as a result of less EC loss. \sa{Similar broadening of individual components of the SED is also observed for all the comparative cases mentioned in table \ref{tab:CD}. As a result, the Compton dominance obtained from the power of each components (CD$^{\rm power}$) suggests to have a higher EC power compared synchrotron power.}
The \sa{CD$^{\rm peak}$ and CD$^{\rm power}$ values} in Ref\_s10\_A case is estimated to be of the order of $\approx$ 0.4 \sa{and 2.5, respectively}.
Also, we noticed that the \sa{CD$^{\rm peak}$ and CD$^{\rm power}$ values} decreases with a decrease in the $\kappa$ value and it is found to be nearly 0.08 \sa{and 1.2, respectively} for Ref\_s10\_B case. \sa{The values obtained from CD$^{\rm power}$ and CD$^{\rm peak}$ seem to follow a similar pattern.}

In summary, the synthetic SEDs generated show a presence of mild Compton dominance \sa{(CD$^{\rm peak}$)}, particularly for the run with lower magnetisation. The small value can primarily be attributed to the fact that the EC emitting region is irradiated by the photon source that essentially lies beneath \sa{(mono-directional photon field)}. \sa{However, if we take into account an isotropic external photon distribution, we may expect to obtain higher CD$^{\rm peak}$ due to the presence more external photons. Furthermore, isotropic electron distribution is a zeroth order assumptions. According to \cite{kelner_2014}, adopting an anisotropic distribution of electrons may have an impact on the measured emission, particularly on the peak of the SED. Implementing such a distribution, though, would be complex and beyond the scope of the current work.}

\section{Summary}\label{sec:summary}

MHD instabilities are one of the vital processes responsible for the observed emission features of blazar jets. The growth of these instabilities lead to the generation of shocks, resulting in the acceleration of particles up to very high energies, which is considered to be one of the prime reasons for the high energy flares seen in these jets.
In this work, we have performed a parametric study characterising the EC process and implemented it in the hybrid framework of the \texttt{PLUTO} code.
Furthermore, we have studied the significance of different radiation mechanisms by performing a numerical simulation of a 2D relativistic slab jet. Lastly, we looked at the multi-wavelength emission of a 3D plasma column in differently magnetised environments where MHD instabilities were present.
The key results of this work are summarised as follows.

\begin{itemize}
    \item We observe a sudden rise in X-ray and optical flux due to the acceleration of emitting electrons near the vicinity of freshly formed shocks generated due to jet instabilities.
    
    \item The impact of such localised shocks is manifested in the SEDs as we find evidence of flattening during the active growth phase of jet instability.
    
    \item Subsequent to the passing of shock, the particles cool towards lower energy and their evolution is dominated by losses due to the EC process for the chosen set of parameters. This transformation of dominance in radiating processes gives rise to a change of slope in the SEDs between the optical and X-ray bands.
    
    \item The effect of radiative cooling due to the interplay of synchrotron and EC processes is clearly reflected by the shifts in the discrete correlation function. In particular, during the shocked phase, X-ray emission leads all other wave-bands. During the later evolutionary stages, correlation is observed between the radio and $\gamma$-ray emissions owing to the up-scattering of external photons via low energy radio electrons giving rise to $\gamma$-ray emission through the EC process.
    
    \item The impact of including the EC process is also evident from the observation of the Compton dominance \sa{(CD$^{\rm peak}$)} of $\sim$\,10 for the case with low magnetisation. Such values of Compton dominance are typically seen in Low Synchrotron Peaked (LSP) blazars.
    
\end{itemize}

In summary, the present work highlights a unique multi-zone and time-dependent framework for blazar that can qualitatively provide insights on the role played by the interplay of instability driven shocks in the presence of two radiative loss mechanisms.
Further, these jets also show a high degree of linear polarisation. Recently, several studies have shown that the observed optical polarisation variability is correlated with high-energy flaring activity. In the subsequent work, we plan to study polarisation signatures of the relativistic  jets by performing 3D simulations that are prone to both axisymmetric and non-axisymmetric instabilities.

\begin{acknowledgements}
      The authors are grateful to Dmitry Khangulyan for the detailed discussion and insightful comments. SA is supported by the DST INSPIRE Fellowship and would like to acknowledge the support for her Ph.D. BV would like to acknowledge the support from the Max Planck Partner Group Award. IKD thank Department of Science and Technology’s Swarnajayanti Fellowship (DST/SJF/PSA-03/2016-17) at IISc Bangalore for financial support. The authors are also grateful to the reviewer for his thorough reading and insightful comments, which helped to improve the manuscript. All the computations presented in this work are carried out using the facilities provided at IIT Indore and the Max Planck Institute for Astronomy clusters: VERA, which is a part of the Max Planck Computing and Data Facility (MPCDF). 
\end{acknowledgements}

\bibliographystyle{aa}
\bibliography{reference}

\appendix

\section{\sa{Validation of implementation of EC}}

\sa{In the following, we have discussed the cooling effect of a single emitting particle due to the radiative processes such as synchrotron and external Compton. Further, in section \ref{sec:sj_spectra}, particle evolution of a single macro-particle is discusses in the context of relativistic slab jet problem. }

\subsection{Calculation of \sa{radiative} cooling time}\label{sec:coolingtime}

\begin{figure}[ht]
    \centering
    \includegraphics[scale=0.22]{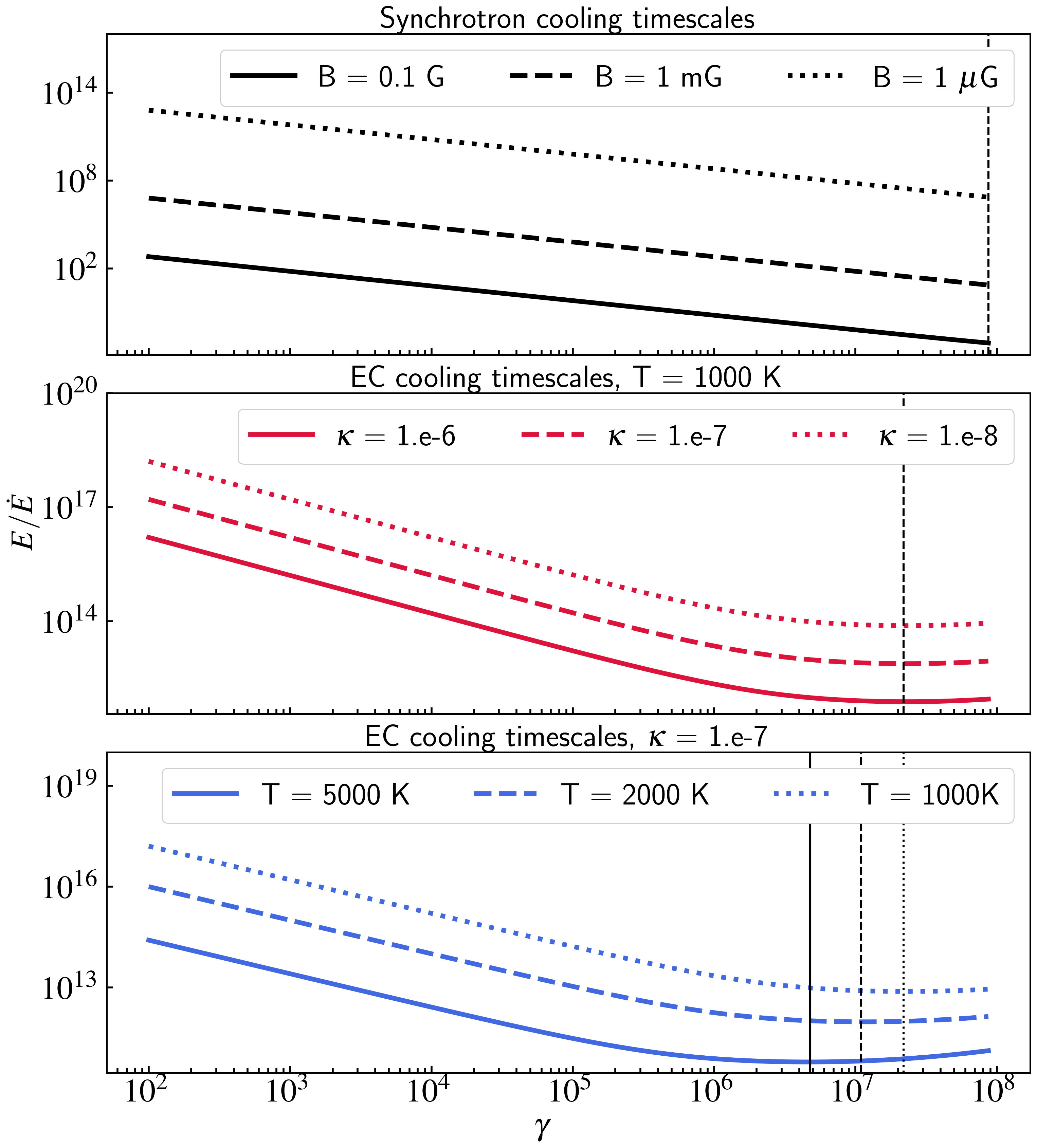}
    \caption{Cooling time scale of the emitting electron for only synchrotron process (top panel) and only EC process (middle and bottom panels). The black dotted lines indicate the energy for which the cooling time is minimum. The different line styles correspond to different parameters as stated in the plot.}
    \label{fig:cool_time}
\end{figure}

Ignoring the loss due to adiabatic expansion, we estimated the cooling time for synchrotron and EC processes for several magnetic field strength, photon field temperature and $\kappa$ values by segregating the each terms of Equation \ref{eq:characteristic} individually. \sa{It should be noted that while estimating the cooling times for various radiation mechanisms, the loss due to adiabatic expansion was purposely turned off. However, while performing the above-mentioned simulations, we took into consideration the loss owing to adiabatic expansion as well as the loss due to synchrotron, IC-CMB, and external Compton.} The top panel of figure \ref{fig:cool_time} represents the cooling time scales for synchrotron processes only, whereas the bottom two panels represent the cooling time scales for solely EC mechanisms. These time-scales are estimated for an initial power law distribution of particles with energy cutoffs $\gamma_{\rm min}$ = 10$^{2}$ and $\gamma_{\rm max}$ = 10$^{8}$ and a power law index $p$ = 6. \sa{Here, we introduced a parameter $\gamma_{\rm cool}$ which is define as a energy for which the radiative cooling timescale is minimum.} The plots of the synchrotron cooling timescale show that the high energy particle cool faster compared to the low energy particle. Additionally, stronger magnetic field strength corresponds to stronger cooling and, therefore, the cooling timescale is much less. However, $\gamma_{\rm cool}$, remains the same for all B-field values.
As seen from the middle and bottom panels, with an increase in T and $\kappa$, the cooling time decreases. The $\gamma_{\rm cool}$ shifts to a lower energy range with increasing value of T, on the other hand, it does not have dependence on the $\kappa$. Such parametrical behaviour of $\gamma_{\rm cool}$ is shown in all panels of figure \ref{fig:cool_time} with black dashed lines. The functional form given in Equation \ref{eq:f(E)_EC} implies that at the lower energy range, the loss rate due to EC is similar to that of synchrotron loss. However, in the higher energy range, it has a logarithmic profile (see Equation 34 of \cite{Khangulyan_2014}). Hence, unlike in synchrotron, where the maximum energy of the electron is responsible for most of the \sa{observed emission or luminosity}, in EC, the emission would be maximum when the energy of an electron is equivalent to the energy of the photon.  Therefore, the cooling time scale would be minimum for an electron, where the transition occurs from Thomson to the Klein Nishina limit. \sa{In other words, the emitting particle that has the minimum cooling time would have the maximum contribution to the observed emission. Furthermore, adopting a particle distribution other than the power-law distribution may alter the cooling timescale profile. However, the typical behaviour of the particle in the cooling process would be the same.}

\subsection{Evolution of particle spectra}\label{sec:sj_spectra}

\begin{figure}[ht]
    \centering
    \includegraphics[scale=0.16]{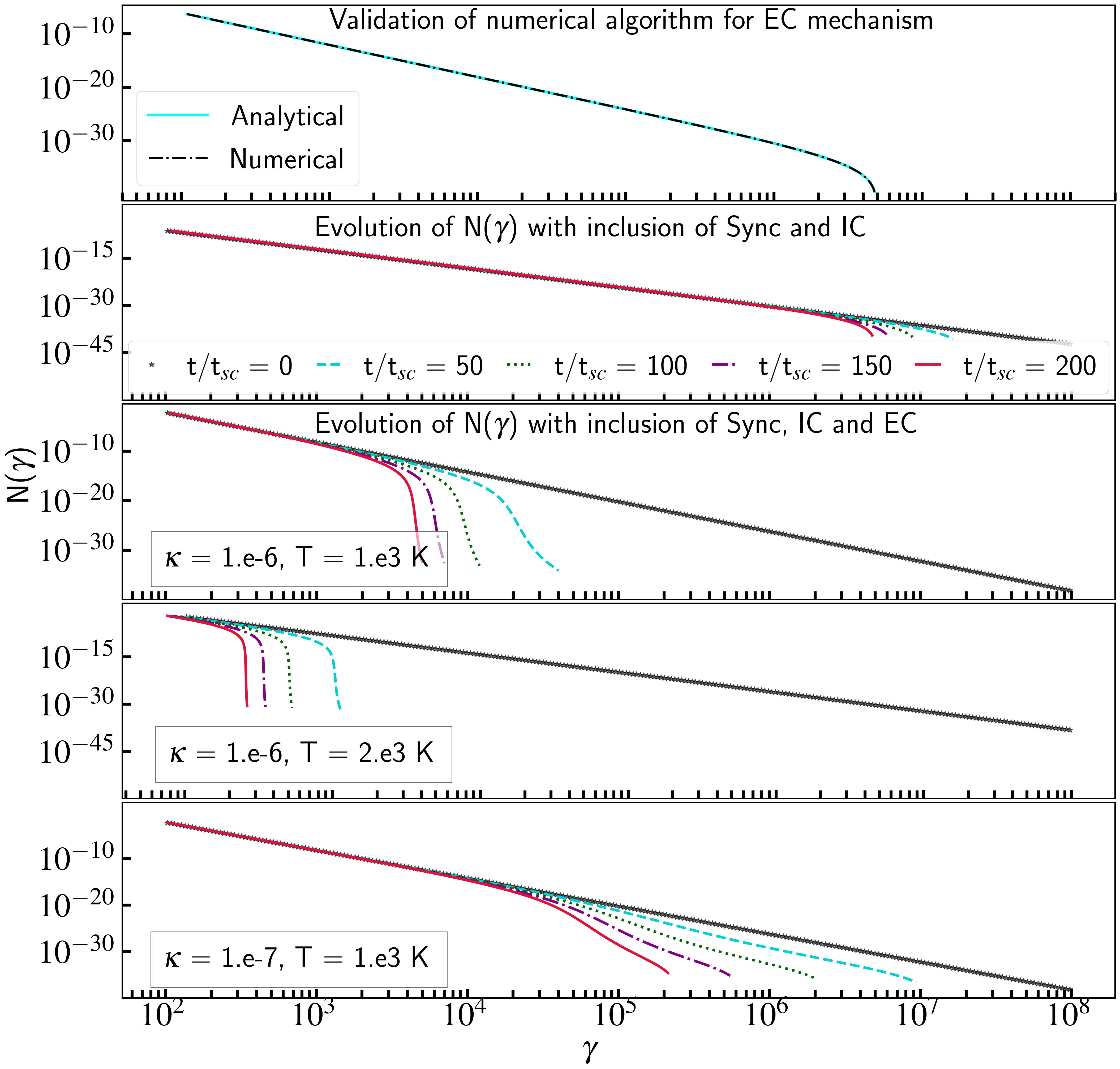}
    \caption{Time evolution of particle spectra for a single test macro-particle \sa{in the context of 2D slab jet problem} with and without inclusion of the EC. The top panel shows the spectra at t/t$_{\rm sc}$ = 200 for the case-1 with both analytical and numerical solution.}
    \label{fig:time_evol_Ne}
\end{figure}

To calculate the energy loss rate due to several radiative processes, \cite{Vaidya_2018} employed an analytical solution to solve Eq. \ref{eq:characteristic} with c$_{3} = 0$. As already mentioned in section \ref{sec:IC_method}, we have included an additional term due to the EC process and solved the Equation \ref{eq:characteristic} numerically by implementing the fourth order Runge-Kutta method (RK4). For validation purposes, we have shown the particle spectra of a single test macro-particle from case-1 \sa{of 2D slab jet problem (see section \ref{sec:SJ_setup&Emis})} at the final time of the simulation in the top panel of figure \ref{fig:time_evol_Ne}. In this panel, the cyan solid line depicts the spectra where Equation \ref{eq:characteristic} is solved analytically and the black dashed line represents the spectra where it is solved using the RK4 method. Note that, in both the scenarios (without the inclusion of the loss term due to the EC process), the spectra show a great degree of overlap. This validates the correctness of extending our analytical method to the numerical approach. This extension is imperative to include the term related to EC losses.

The energy-loss rate of an electron with and without the inclusion of EC differs significantly from each other. The effect of such energy loss is distinctly visible in figure \ref{fig:time_evol_Ne}, which exhibits the time evolution of particle spectra for a single macro-particle. In our simulations, a single macro-particle refers to a collection of a number of micro-particles or leptons. The second panel of figure \ref{fig:time_evol_Ne} shows time-evolving particle spectra for a single test macro-particle incorporating loss due to only the synchrotron and IC-CMB mechanisms. The next three panels at the bottom manifest the spectra of the single macro-particle with the incorporation of energy loss due to the EC process, in addition to synchrotron and IC-CMB for different parameters. \sa{To understand the dependence of different parameters responsible for the EC process, a set of values for T (= 1000 K and 2000 K) \citep{Donea_2003_torusTemp, Malmrose_2011, Oyabu_2017_torusTemp} and $\kappa$ (= 10$^{-7}$ and 10$^{-6}$) are adopted. }
A noticeable difference is visible between the particle spectra shown in the second and bottom three panels of figure \ref{fig:time_evol_Ne}. The addition of EC loss in Equation \ref{eq:characteristic} results in the cooling of electron faster and stronger compared to the case that includes loss due to synchrotron and IC-CMB only. In addition, higher temperature and $\kappa$ correspond to stronger loss, and therefore, the $\gamma_{\rm max}$ of the emitting particle becomes very small. A similar effect \sa{was also} seen while considering the behaviour of a collection of macro-particles randomly distributed throughout the jet \sa{(section \ref{sec:sj_emission})}.

\end{document}